\documentclass[11pt]{article}

\usepackage{amsmath,amsthm,amsfonts,amssymb,color,hyperref,anysize,enumitem,graphicx,epstopdf,algorithm,algpseudocode,cleveref,multirow}
\usepackage[usenames,dvipsnames]{xcolor}
\usepackage[margin=0.6in]{geometry}

\newcommand{\rr}{\mathbb{R}}
\newcommand{\rrplus}{\rr^+}
\newcommand{\nn}{\mathbb{N}}

\newcommand{\ra}{\rightarrow}

\newcommand{\vdd}{\vspace*{0.05in}}

\DeclareMathOperator*{\argmin}{arg\,min}

\newenvironment{pf}{\begin{proof}[\emph{\textbf{Proof: }}]}{\end{proof}}

\newenvironment{pfof}[1]{\begin{proof}[\emph{\textbf{Proof of #1: }}]}{\end{proof}}

\usepackage{framed}

\newcommand{\boxdef}[1]{
\begin{framed}
\setlength{\topsep}{1pt}
\begin{definition}
\normalfont #1
\end{definition}
\end{framed}
}

\newcommand{\boxthm}[1]{
\begin{framed}
\setlength{\topsep}{1pt}
\begin{theorem}
\normalfont #1
\end{theorem}
\end{framed}
}

\newcommand{\boxlem}[1]{
\begin{framed}
\setlength{\topsep}{1pt}
\begin{lemma}
\normalfont #1
\end{lemma}
\end{framed}
}

\newcommand{\boxprop}[1]{
\begin{framed}
\setlength{\topsep}{1pt}
\begin{prop}
\normalfont #1
\end{prop}
\end{framed}
}

\usepackage{tcolorbox}
\tcbuselibrary{theorems}

\newtcbtheorem[auto counter]{coloredDEF}{Definition}
{colback=blue!5,colframe=blue!40!black,fonttitle=\bfseries,before={\vspace{0.3cm}},after={\vspace{0.3cm}}}{de}

\newtcbtheorem[auto counter]{coloredNOT}{Notation}
{colback=blue!5,colframe=blue!40!black,fonttitle=\bfseries,before={\vspace{0.3cm}},after={\vspace{0.3cm}}}{no}

\newtcbtheorem[auto counter]{assume}{Assumption}
{colback=green!5,colframe=green!60!black!80,fonttitle=\bfseries,before={\vspace{0.3cm}},after={\vspace{0.3cm}}}{assume}

\newtcbtheorem[auto counter]{coloredLEM}{Lemma}
{colback=red!5,colframe=red!60!black,fonttitle=\bfseries,before={\vspace{0.3cm}},after={\vspace{0.3cm}}}{lm}

\newtcbtheorem[auto counter]{coloredTHM}{Theorem}
{colback=purple!5,colframe=purple!50!black,fonttitle=\bfseries,before={\vspace{0.3cm}},after={\vspace{0.3cm}}}{th}

\newtcbtheorem[auto counter]{coloredPROP}{Proposition}
{colback=purple!5,colframe=purple!50!black,fonttitle=\bfseries,before={\vspace{0.3cm}},after={\vspace{0.3cm}}}{pr}

\newtcbtheorem[auto counter]{coloredCOR}{Corollary}
{colback=purple!5,colframe=purple!50!black,fonttitle=\bfseries,before={\vspace{0.3cm}},after={\vspace{0.3cm}}}{co}

\newtheorem{theorem}{Theorem}[section]
\newtheorem{lemma}[theorem]{Lemma}
\newtheorem{corollary}[theorem]{Corollary}

\newtheorem{prop}[theorem]{Proposition}
\newtheorem{definition}[theorem]{Definition}
\newtheorem{shortexercise}[theorem]{Short Exercise}
\newtheorem{exercise}[theorem]{Exercise}

\newcommand{\calF}{\mathcal{F}}

\newcommand{\calI}{\mathcal{I}}

\newcommand{\calU}{\mathcal{U}}

\newcommand{\bbb}{\mathbf{b}}

\newcommand{\bbn}{\mathbf{n}}

\newcommand{\bbp}{\mathbf{p}}

\newcommand{\bbu}{\mathbf{u}}

\newcommand{\bbz}{\mathbf{z}}

\newcommand{\vuppp}{\vspace*{-0.06in}}

\usepackage{mathtools}
\usepackage{cleveref}

\def\p{\mathbf{p}}

\def\q{\mathbf{q}}
\def\x{\mathbf{x}}
\def\y{\mathbf{y}}
\def\z{\mathbf{z}}
\def\v{v}
\def\q{q}
\def\s{\phi}
\def\Bphi{\boldsymbol{\phi}}

\def\X{\mathbf{X}}

\DeclarePairedDelimiterX{\infdivx}[2]{(}{)}{%
	#1\;\delimsize\|\;#2%
}
\DeclarePairedDelimiterX{\innerProd}[2]{\langle}{\rangle}{%
	#1 , #2%
}

\title{Stability and Efficiency of Personalised Cultural Markets}
\author{Haiqing Zhu\vdd\\
\small Australian National University\\
\small Canberra, Australia\\
\small \texttt{haiqing.zhu@anu.edu.au}
\and
Yun Kuen Cheung\vdd\\
\small Royal Holloway University of London\\
\small Egham, UK\\
\small \texttt{yunkuen.cheung@rhul.ac.uk}
\and
Lexing Xie\vdd\\
\small Australian National University\\
\small Canberra, Australia\\
\small \texttt{lexing.xie@anu.edu.au}
}
\date{}

\begin{document}

\maketitle

\begin{abstract}
This work is concerned with the dynamics of online cultural markets,
namely, attention allocation of many users on a set of digital goods with infinite supply. 
Such dynamics are important in shaping processes and outcomes in society,
from trending items in entertainment, collective knowledge creation, to election outcomes.
The outcomes of online cultural markets are susceptible to intricate social influence dynamics,
particularly so when the community comprises consumers with heterogeneous interests.
This has made formal analysis of these markets improbable.
In this paper, we remedy this by establishing robust connections between influence dynamics and optimization processes,
in trial-offer markets where the consumer preferences are modelled by multinomial logit.
Among other results, we show that the proportional-response-esque influence dynamic is equivalent to stochastic mirror descent on a convex objective function,
thus leading to a stable and predictable outcome.
When all consumers are homogeneous, the objective function has a natural interpretation as a weighted sum of
efficiency and diversity of the culture market. 
In simulations driven by real-world preferences collected from a large-scale recommender system,
we observe that ranking strategies aligned with the underlying heterogeneous preferences are more stable,
and achieves higher efficiency and diversity.
More broadly, we see this work as the first step in connecting 
computational methods for classical markets to the problem area of online attention and recommender systems.
We hope this result paves the way to posing and answering a diverse set of research questions in this area.
\end{abstract}

\section{Introduction}

Online content platforms are major sources for everyday entertainment, and the attention allocation within a platform provides a market setting of interest. 
There are strong social and economic motivations for the stakeholders to understand these markets.
The platform providers ask how they can improve user experience and raise profits.
They may also want to enforce diversity of the cultural products, so as to achieve a sustainable business model.
The content producers are interested in strategies to improve their products, so as to gain more popularity and raise revenues.
Regulatory bodies and the user population seek to understand the market dynamic and its drivers,
with the goal of making attention markets more transparent, accountable and fair.

However, understanding such markets is challenging, since their outcomes are susceptible to intricate social influence dynamics among customers,
which in turn are affected by the recommender systems of the platform providers.
While various analyses have been done for the classical markets (e.g., Arrow-Debreu/exchange markets, Fisher markets),
the online cultural markets (or more generally, \emph{attention economies}) have several key aspects different from the classical markets,
so it is not clear if known analyses directly apply.
In classical markets, goods are scarce. Their prices act as the coordination signals to balance demand and supply.
Typically, there exist moderate prices that lead to equilibrium. In online cultural markets, however,
the digital goods can be reproduced with essentially no cost, so their supplies are unlimited.
Users' attention is the scarce commodity that the producers compete for.
Typically, the influence dynamics and recommender systems tend to \emph{cascade}, i.e., to promote goods which already got high attentions.
This is known to lead to polarizing and unpredictable outcomes.

Since the seminal empirical work by Salganik et al.~\cite{salganik2006ExperimentalSO}, now dubbed \textsc{MusicLab},
several mathematical models have been proposed to describe it~\cite{krumme2012QuantifyingSI,abeliuk2015benefits,maldonado2018popularity},
but all of them assume that user preferences are homogeneous.
This is in stark contrast to the rich literature and wide-spread practice of recommender systems that are focused on estimating
and catering to heterogeneous preferences in user populations. 
On the other hand, recent work in classical markets (esp Fisher markets) offer a range of results to understand equilibria
from an algorithmic and optimization perspective~\cite{Zhang2011PR,birnbaum2011distributed,cheung2018dynamics}.
One may wonder: can recent results on Fisher markets be extended to describe the implicit computations in cultural markets?
Specifically for the \textsc{MusicLab} model~\cite{krumme2012QuantifyingSI}, customer behaviors are characterized by a two-step trial-offer (T-O) process:
first, they select a product to try; second, they decide to purchase or not.
The first step is a stochastic process where the randomness depends on the intrinsic \emph{appeal} of the products,
and also the history of past purchases by customers, creating a feedback loop.
The second step is random depending on the intrinsic \emph{quality} to the customer.
\begin{figure}[bt]
\centering
\includegraphics[width=0.85\textwidth]{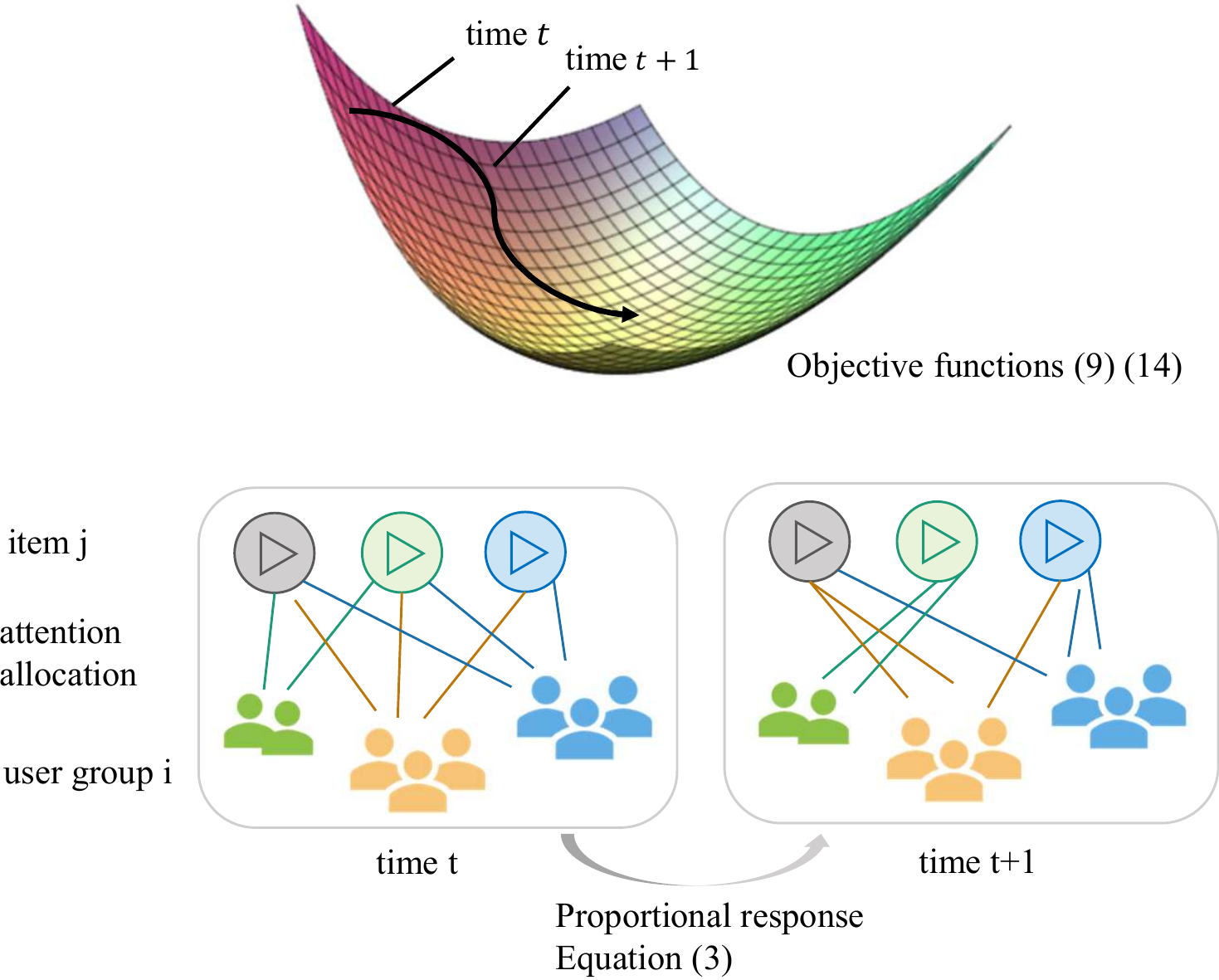}
\caption{A core contribution of this paper is to provide an optimization view (Top) of cultural markets (Bottom), which affords new results on stability, efficiency and equilibrium behavior. (Bottom) An illustration of a cultural market with several types of users interacting with a few items (color similarity between users and items indicate differing matches in preferences). Users allocate attention to the items based on proportional-response-esque dynamic (see \Cref{choice TO hetero}). }
\label{fig:bigalpha}
\end{figure}

\subsection*{Our contributions}

The main themes of this work are to establish robust connections between stochastic T-O markets and optimization,
and to use these connections to rigorously show that the influence dynamics in these markets are stable.
For the homogeneous markets, we discover two objective functions for which the equilibrium of a T-O market
is maximiser of these objectives. The first objective is the ``total utility'' of the market.
The second objective is of particular interest due to its natural interpretation.
It is a weighted sum of the efficiency and the diversity of the market shares in the market, as measured by the Shannon entropy.
While efficiency is a natural benchmark, diversity in cultural market is also important for the healthy development of the platform.
The diversity not only broaden the customer base, it also provides financial support to the less popular producers to keep them in the cultural industry.
Thus, it is of the platform providers' interest to strike for a balance between efficiency and diversity.

Interestingly, we show that the influence dynamic is indeed equivalent to stochastic mirror descent on the second objective.
This suggests the dynamic is implicitly optimizing the natural objective in the market.
A significant consequence is this allows us to present a new proof of a result of Maldonado et al.~\cite{maldonado2018popularity}
that the dynamic converges to an equilibrium of the market almost surely.

For heterogeneous markets, we show that the equilibrium is optimizing an \emph{ex-post} version of Nash social welfare.
In classical Fisher market, Nash social welfare is the product of users' utilities, whereas each utility is raised to a power of the user's budget.
In our case, the power is the budget times the \emph{efficiency} for that user at the equilibrium.
Then we turn our focus to two interesting sub-classes of heterogeneous markets, namely
(i) the users have the same appeals on the items, but they perceive the qualities of the items differently;
(ii) the users perceive same qualities on the items, but they have different appeals on the items.
For (i), we show that it is equivalent to a homogeneous market.
For (ii), we design a new objective function, where the influence dynamic is equivalent to stochastic mirror descent on the objective.
Again, this allows us to show the dynamic converges to an equilibrium almost surely.

The robust connection between the dynamics and optimization processes echoes with \emph{the self-reinforced efficiency} of some economic systems,
for which there exist natural dynamics or algorithms that can attain equilibrium,
while the equilibrium optimizes popular efficiency measure like social welfare.
See Related work below for more relevant discussions.

We perform simulations using user preferences from the well-known MovieLens-100K dataset~\cite{harper2015movielens}.
We observe that accounting for heterogeneous user preferences improves efficiency in cultural markets while preserving stability.
We examine the (user-centric) efficiency and (item- or producer- centric) diversity measures across three ranking strategies: 
random, quality-driven, and popularity-driven.
Results confirm quality ranking being more efficient and more diverse than popularity ranking
which was implemented in \textsc{MusicLab}~\cite{salganik2006ExperimentalSO} and is known to be unstable.

The rest of this paper is organized as follows. After describing the models in \Cref{sec:model}, we present our main results formally in \Cref{sec:results}.
In \Cref{sect:analysis}, we provide an overview of the techniques we employ for proving the main results.
This is followed by a discussion of empirical observations in \Cref{sect:experiments}.

\subsection*{Related work}

As early as in 1971, Simon~\cite{simon1971designing} pointed out that in an information-rich world, attention becomes the new scarcity that information consumes.
Examples of attention economies include entertainment such as music, film and television~\cite{salganik2006ExperimentalSO,bell2007lessons},
political campaigns and votes~\cite{BondFJKMSF2012}, scientific publications and researchers~\cite{Fortunato2018TheSO}.
Since Simon's visionary statement, the research community has formulated economic question about attention in a number of different ways,
such as articulating the phenomenon of attention scarcity in corporate life~\cite{davenport2001attention},
diagnostic criteria for attention scarcity and solving it as (one off) allocation problems~\cite{falkinger2007attention,falkinger2008limited},
or connecting attention allocation to advertising revenue~\cite{evans2020economics}.
A recent study by Vosoughi et al.~\cite{VosoughiFalseNews2018} showed that false news spreads faster online, suggesting that besides quality,
appeal (e.g., novelty of the false news and the emotion it stimulates) of a digital good is crucial in social influence.
More broadly, the web research community have measured attention to items by individual users~\cite{tong2020brain},
a large set of users~\cite{wu2018beyond}, and attention among a network of items~\cite{Wu2019EstimatingAF}.

The concept of self-reinforced efficiency of economic systems can be traced back to the ``invisible hand'' metaphor of Adam Smith.
One of the first analytical confirmations of the concept is the famous First Welfare Theorem,
which states that a market equilibrium of any complete market is Pareto efficient~\cite{Lange1942,Arrow1951,Debreu1959}.
Furthermore, in a broad class of markets called Eisenberg-Gale markets,
market equilibrium optimizes a popular efficiency measure called Nash social welfare \cite{EG59,Eisenberg61,JainVazirani-EG2010}.
On the other hand, in combinatorial auction, any Walrasian equilibrium (if exists) optimizes the social welfare \cite{BM1997}.
In many of these economic systems, there are natural adaptive price/bidding dynamics
(e.g., t\^atonnement~\cite{walras,ABH59,CMV05,CF08,CCR12,CheungCD2020,Cheung2014,CheungHN19}, proportional response \cite{WZ2007,LLSB08,Zhang2011PR,birnbaum2011distributed,cheung2018dynamics,BMN2018,CheungHN19,GaoKroer2020,BranzeiDR2021,CheungLP2021,CheungLSP2022})
or auction algorithms (e.g., ascending-price auctions~\cite{KelsoCrawford1982,NS2006}) that attain efficient equilibria.
As we shall see, the influence dynamics we study are indeed a stochastic version of proportional response.
\section{Model: The Trial-Offer Market with Heterogeneous User Types}\label{sec:model}

First, we describe stochastic trial-offer (T-O) market, in which users come to the platform one-by-one to try and purchase the items.
We introduce measures of efficiency and diversity. Then we describe a continuous and deterministic analogue of the stochastic model which will be useful for analysis.
In this work, we use {\em purchase} to denote user completing a transaction on an item, where the resource a user {\em spends} is attention.
One may think of it as a unit amount of time. Without loss of generality,
we assume that each user has the same {\em budget} of attention, and that each item {\em costs} one unit of attention.
This model generalises cultural markets specified by Krumme et al.~\cite{krumme2012QuantifyingSI}
and Maldonado et al.~\cite{maldonado2018popularity} to heterogeneous types of users.

\subsection{Stochastic Trial-offer (T-O) Market}

Let $\calU$ denote the types of users and $\calI$ denote the set of items.
The fraction of Type-$i$ users is denoted by $w_i$; note that $\sum_{i=1}^{|\calU|} w_i = 1$.
If $|\calU|=1$, we say the market is \emph{homogeneous}, otherwise it is \emph{heterogeneous}.

The dynamic starts at time $t=0$. At each time $t\ge 1$, a random user comes to the platform and tries an item, and then she decides to purchase the item or not.
Let $d_j^t$ denote the number of purchases of item $j$ up to time $t$. To ensure that all items have a positive probability to be tried in the initial rounds,
we assume that each item was purchased at least once before the dynamic starts, i.e., $d_j^0 \ge 1$ for every item $j$.
The \emph{market share} of item $j$ at time $t$ is simply the fraction of all purchases that goes to item $j$:
\[
\phi^t_j ~:=~ \frac{d^t_j}{\sum_{k=1}^{|\calI|} d^t_k}.
\]
The possible market shares lie on a simplex, denoted by $\Delta$:
\[
\Delta = \left\{\Bphi\in \mathbb{R}^{|\calI|} : \sum_{j=1}^{|\calI|} \phi_j = 1, \phi_j\geq 0 \text{ for any item $j$}\right\}.
\]
If the user at round $t$ is of Type-$i$, the probability that she will try item $j$ is modelled as a multinomial logit,
a common type of discrete choice model~\cite{greene2017econometric}:
\begin{equation}
\tilde y_{ij}^t = \frac{v_{ij} (\phi^{t-1}_j)^{r_i}}{\sum_{k=1}^{|\calI|} v_{ik} (\phi^{t-1}_k)^{r_i}}~,\label{eq:mnlchoice}
\end{equation}
$v_{ij} \ge 0$ is a parameter that depicts the \emph{visibility} of item $j$ to Type-$i$ users, which depends on the \emph{appeal} of the item itself,
and also how the item is promoted or ranked with respect to other items.
$r_i > 0$ is a parameter called \emph{feedback exponent}, which depicts the strength of feedback signal for Type-$i$ users.\footnote{$r_i=0$
means no {\em social} feedback signal from the current market share, whereas $r_i \rightarrow \infty$ means only the most popular item will be chosen in the next round.
If the denominator $\sum_{k=1}^{|\calI|} v_{ik} (\phi^{t-1}_k)^{r_i} = 0$, then this probability is defined as $0$.}

After a Type-$i$ user tries an item $j$, they purchase the item with probability $q_{ij} \in [0,1]$,
which intuitively reflects the {\it quality} of an item that may be unknown before it is tried.

In a homogeneous market, there is only one user type, we will drop all indices $i$ from the notations, resulting in $v_j, q_j, r$, and clearly $w_1=1$.

The dynamic is the result of two interacting factors. The first is the user-specific visibility and quality factors $v_{ij}$ and $q_{ij}$,
which generalise recent models that analyse homogeneous attention markets with feedback loops \cite{krumme2012QuantifyingSI,jiang2019degenerate}. 
The second is a social feedback signal  $(\phi_j^{t-1})^{r_i}$ based on the overall popularity of the item,
such as the one implemented by the original \textsc{MusicLab} experiment~\cite{salganik2006ExperimentalSO}, or the number of downloads and likes on myriads of internet platforms.
This feedback dynamic is also similar to proportional response in Fisher markets~\cite{cheung2018dynamics}, which we will exploit for obtaining key results in \Cref{sec:results}.

Ranked list is one of the most popular forms of presenting a set of items to users,
and a salient factor affecting the visibility of an item is its position in such a list~\cite{craswell2008experimental}.
If the positions are fixed throughout the attention dynamic, $v_{ij}$ remains constant. Our theoretical results focus on this case.
In \Cref{sect:experiments}, we empirically explore how strategies of dynamically positioning the items by the platform will affect the outcome. 
We compute the probability of item $j$ being the next purchase by manipulating the trial and purchase probabilities.

\boxlem{\label{lem:market-eff}
In a stochastic T-O market defined above,
the probability that the next purchase is for item $j$, denoted by $p_j(\Bphi)$,
is a function of the current market share $\Bphi$, 
given by $p_j(\Bphi) = y_j(\Bphi) / (\sum_{k=1}^{|\calI|} y_k(\Bphi))$, where
\begin{equation}
y_j(\Bphi) ~:=~ \sum_{i=1}^{|\calU|} w_i q_{ij} \cdot \frac{v_{ij} (\phi_j)^{r_i}}{\sum_{k=1}^{|\calI|} v_{ik} (\phi_k)^{r_i}}.\label{choice TO hetero}
\end{equation}
$y_j(\Bphi)$ represents the probability that item $j$ is tried and then purchased by any user group.
In particular, for the homogeneous case, the probability that the next purchase is for item $j$ is
\begin{equation}
\frac{v_j q_j (\phi_j)^{r}}{\sum_{k=1}^{|\calI|} v_k q_k (\phi_k)^{r}}.\label{choice TO} \vuppp\vuppp
\end{equation}
}

\subsection{Trial-Offer Market Equilibrium}

For $\Bphi$ to be a stationary point in this stochastic process, it must satisfy $p_j(\Bphi) = \phi_j$
for all items $j$. This motivates the following equilibrium notion.

\boxdef{\label{defn:TOME}
For any T-O market, we say a market share $\Bphi$ is a trial-offer market equilibrium (TOME) if $\p(\Bphi) = \Bphi$.
We say $\Bphi$ is an interior TOME if it is a TOME with $\phi_j > 0$ for all items $j$.}

The following theorem establishes that TOME exists in heterogeneous T-O market under mild conditions.
It extends previous results on homogeneous markets~\cite{maldonado2018popularity}.
The proof, which uses the Brouwer's fixed-point theorem, is presented in \Cref{app:property}.

\boxthm{[Existence of TOME.]\label{thm:exist-TOME}
If for any user type $i$ the population fraction $w_i > 0$ and the feedback exponent $0 < r_i < 1$, 
then the T-O market must have a TOME $\Bphi^*$, in which $\phi_j^* > 0$ for any item $j$ with $v_{ij} q_{ij} > 0$ for at least one user type $i$.}

If there is an item $j$ with $v_{ij} q_{ij} = 0$ for all user types $i$, then its equilibrium market share must be zero
 -- leading to $\Bphi^*$ on the relative boundary of the simplex
 -- so we may ignore the item in analysis. In the rest of this paper, we assume there is no such item in the markets.

In the homogeneous case, we can explicitly compute the unique TOME $\Bphi^*$ if $0<r<1$:
\begin{equation}
\phi^*_j ~=~ \frac{(v_j q_j)^{1/(1-r)}}{\sum_{k=1}^{|\calI|} (v_k q_k)^{1/(1-r)}},\label{eq:TOMEhomo}
\end{equation}
provided that $v_k q_k > 0$ for some item $k$. It is easy to verify $\Bphi^*$ is a stationary point by plugging it into \Cref{defn:TOME}.
\Cref{sec:results} specifies how to obtain $\Bphi^*$ and argues for its uniqueness in the interior of the simplex $\Delta$.

\subsection{Efficiency and Diversity Measures}

An online platform may be interested in maximising the probability of successful transaction among all items, which is $\sum_{j=1}^{|\calI|} y_j(\Bphi)$. 

\boxdef{\label{defn:efficiency}
Given market share $\Bphi\in \Delta$, define the T-O market efficiency as $E(\Bphi) := \sum_{j=1}^{|\calI|} y_j(\Bphi)$.}

A platform may also be interested in promoting diversity among items.
A natural measure of diversity is the \emph{Shannon entropy}, which is the standard measure of uncertainty of a probability distribution in information theory \cite{Cover2006}.
Given market share $\Bphi\in \Delta$, its Shannon entropy is
\begin{equation}
H(\Bphi) = - \sum_{j=1}^{|\calI|} \phi_j \log \phi_j.\label{eq:entropy}
\end{equation}

\subsection{The Deterministic T-O Market Dynamic}

This model is analogous to the stochastic T-O model. It will be useful for analysis. There is one user of each type $i$, whose budget is $w_i$.
The budget $w_i$ corresponds to the maximum amount of attention the buyer can afford in the platform.
At each time $t\ge 0$, each buyer $i$ spends an amount of $b_{ij}^t$ for item $j$, subject to the budget constraint $\sum_{j=1}^{|\calI|} b_{ij}^t \le w_i$.
Let the total spending on item $j$ be $b_j^t := \sum_{i=1}^{|\calU|} b_{ij}^t$.
The market share of item $j$ at time $t$ is $\phi_j^t := b_j^t / (\sum_{k=1}^{|\calI|} b_k^t)$.
For $t\ge 1$, the update rule is
\begin{equation}
b_{ij}^t = w_i q_{ij} \cdot \frac{v_{ij} (\phi^{t-1}_j)^{r_i}}{\sum_{k=1}^{|\calI|} v_{ik} (\phi^{t-1}_k)^{r_i}}
~=~ w_i q_{ij} \cdot \frac{v_{ij} (b^{t-1}_j)^{r_i}}{\sum_{k=1}^{|\calI|} v_{ik} (b^{t-1}_k)^{r_i}}. \label{cont dynamic}
\end{equation}

Crucially, there is a natural correspondence between TOME of a stochastic T-O market and the fixed point of the dynamic \eqref{cont dynamic},
summarized by the lemma below. Its proof can be found in \Cref{app:property}.

\boxlem{\label{lem:corr-cont-stochastic}
(i) If $\Bphi^* = (\phi_j^*)$ is a TOME of the stochastic T-O market, set $b_{ij}^* := w_i q_{ij} \cdot \frac{v_{ij} (\phi^*_j)^{r_i}}{\sum_{k=1}^{|\calI|} v_{ik} (\phi^*_k)^{r_i}}$.
Then $\mathbf{b}^* = (b_{ij}^*)$ is a fixed point of the dynamic \eqref{cont dynamic}.

\noindent (ii) If $\mathbf{b}^* = (b_{ij}^*)$ is a fixed point of the dynamic \eqref{cont dynamic}, 
set $b_j^* := \sum_{i=1}^{|\calU|} b_{ij}^*$, then set $\phi_j^* := b_j^* / (\sum_{k=1}^{|\calI|} b_k^*)$.
Then $\Bphi^* = (\phi_j^*)$ is a TOME of the corresponding stochastic T-O market.}

\subsection{Comparison with Classical Fisher Market and Proportional Response}

For readers who are familiar with Fisher market dynamics, the deterministic T-O market dynamic \eqref{cont dynamic}
is reminiscent of the proportional response (PR) dynamic in Fisher markets~\cite{cheung2018dynamics}.
There is one crucial difference though. PR dynamic in Fisher market is same as dynamic \eqref{cont dynamic} but with $b_k^{t-1}$ on the RHS replaced by $b_{ik}^{t-1}/b_k^{t-1}$.
The term $b_k^{t-1}$ in Fisher market is viewed as the \emph{price} of item $k$; a higher price in PR drives down spendings on that item from the buyers.
In contrast, a higher value of $b_k^{t-1}$ in \eqref{cont dynamic}, which corresponds to receiving more attention in the T-O market, will lead to more spending on that item.
This reflects the tendency of cascading in online attention dynamics. 
\section{Results}\label{sec:results}

This section provides an overview of our key new results.
\Cref{ssec:maxutil} establishes a novel connection between TOME in homogeneous markets and two convex objectives.
\Cref{ssec:mirror_descent} shows that update steps in deterministic T-O markets are mirror descent steps for one of the objectives. 
\Cref{ssec:heteroRMA} presents the objective functions for heterogeneous markets.

\subsection{TOME maximises regularised utilities}\label{ssec:maxutil}

First, we establish a robust connection between TOME of {\it homogeneous} T-O market and optimization.
For notational simplicity, let $\bar{q}_j = q_jv_j$, noting that the quality and visibility factors $q_jv_j$ are coupled in both \eqref{choice TO} and \eqref{eq:TOMEhomo}.
We consider the following two constrained optimization problems:
\begin{equation}
	\begin{aligned}
		&\max \quad \sum_{j=1}^{|\calI|} \bar{q}_j\phi_j^r, \\
		&\text{subject to } \Bphi\in\Delta. 
	\end{aligned}\label{TotalUtilityProblem}
\end{equation}
and
\begin{equation}
\begin{aligned}
&\max \quad \Psi(\Bphi) := \sum_{j=1}^{|\calI|} \left(\phi_j \log \bar{q}_j - (1-r) \phi_j \log \phi_j\right), \\
&\text{subject to } \Bphi\in\Delta. 
\end{aligned}\label{Efficiency Entropy Problem}
\end{equation}

We establish the equivalence between the equilibria and the maximisers of the above problems in the following theorem.
The proofs for both simply invoke Lagrangian multipliers, and they are presented in \Cref{app:homo}.

\boxthm{\label{thm:social-welfare}
If $\Bphi^*$ is a maximiser of problem \eqref{TotalUtilityProblem} or problem \eqref{Efficiency Entropy Problem}, then it is a TOME.}

We can view the objective function \eqref{TotalUtilityProblem} as the ``total utility'' since
the choice probability of item $j$ is proportional to the ``utility'' associated with it, which is $\bar{q}_j\phi_j^r$. 
The objective function \eqref{Efficiency Entropy Problem} can be decomposed into two sums, namely
$\sum_{j=1}^{|\calI|} \phi_j \log \bar{q}_j$ and $(1-r) \sum_{j=1}^{|\calI|} -\phi_j \log \phi_j$.
The first sum can be viewed as an alternative measure of total utility,
with the utility of item $j$ being $\log \bar q_j$ weighted by its market share $\phi_j$. 
The second sum is $(1-r)$ times the Shannon entropy of the market share.
When $r=1$, the entropy term disappears, so the optimization problem \eqref{Efficiency Entropy Problem}
becomes trivial: the optimal solution is by setting $\phi_j = 1$ for the highest-utility item $j = \arg\max_k \bar{q}_k$.
As $r$ decreases from $1$, i.e., the strength of feedback signal reduces, the entropy term becomes more significant,
which encourages diversity in the optimal solution.

For $0<r<1$, the objective functions in \eqref{TotalUtilityProblem} and \eqref{Efficiency Entropy Problem} are both strictly concave in $\Bphi$, therefore having a unique maximum.
A crucial advantage of \eqref{Efficiency Entropy Problem} over \eqref{TotalUtilityProblem} is that
mirror descent on \eqref{Efficiency Entropy Problem} provides insight into the convergence
of the stochastic influence dynamics in T-O market as specified in \eqref{eq:mnlchoice} and \eqref{choice TO}.
To formally describe this discovery, we need several concepts in optimization theory, which are discussed next. 

\subsection{T-O update as mirror descent, and TOME convergence for homogeneous markets}\label{ssec:mirror_descent}

\paragraph{Background: Bregman Divergence and Mirror Descent.}
Consider a general constrained convex optimization problem of minimizing a smooth convex function $f(x)$, subject to the constraint $x\in C$ for some compact and convex set $C$.

\boxdef{\label{defn::Bregmann divergence}
Let $C$ be a compact and convex set, and let $h$ be a differentiable convex function on $C$.
The \emph{Bregman divergence} w.r.t.~$h$, denoted by $d_h$, is defined as
\[
d_h(x,y) = h(x) - h(y) - \langle \nabla h(y)~,~ x-y \rangle,
\]
for any $x\in C$ and $y\in \textsf{rint}(C)$.}

The widely used Kullback–Leibler (KL) divergence is a special case of Bregman divergence,
generated by the function $h(x) = \sum_j (x_j \log x_j - x_j)$.

Given a Bregman divergence $d_h$, the corresponding mirror descent update rule is
\begin{equation}\label{general MD rule}
x^{t} ~=~ \argmin_{x\in C} \left\{ \langle \nabla f(x^{t-1})~,~x-x^{t-1} \rangle + \frac{1}{\alpha} \cdot d_h(x,x^{t-1}) \right\},    
\end{equation}
where $\alpha$ is considered as the step-size of the update rule, which may depend on $t$ in general.

\paragraph{New Result: T-O update as Mirror Descent} A key conceptual message of this paper is the equivalence of influence dynamic and mirror descent.
To illuminate this, we first focus on deterministic and homogeneous T-O market.

\boxlem{\label{lem:mirror-descent-homo}
Let $\Bphi^t$ be market share at time $t$ in a homogeneous T-O market, and function $\bbp(\Bphi)$ as defined in \eqref{choice TO}. The update rule 
\begin{equation}\label{deterministic update rule homogeneous}
\Bphi^{t} = \bbp(\Bphi^{t-1})    
\end{equation}
is equivalent to the mirror descent update rule \eqref{general MD rule}
for the optimization problem \eqref{Efficiency Entropy Problem}, in which $d_h$ is taken as the KL divergence,
and $\alpha = 1$.}

The proof of the above lemma is presented in \Cref{app:homo}. Once the equivalence is established,
the convergence to TOME of the deterministic dynamic \eqref{deterministic update rule homogeneous} becomes intuitive;
we will provide the formal argument in \Cref{sect:analysis}. To show that the convergence extends to the stochastic setting,
we follow Maldonado et al.~\cite{maldonado2018popularity} to rewrite the stochastic influence dynamic
as a Robbins-Monro algorithm (RMA) of the deterministic dynamic \eqref{deterministic update rule homogeneous}.
Precisely, the RMA is in the form of $\Bphi^t = \Bphi^{t-1} + \frac{1}{t}\cdot \bbu^t$,
where $\bbu^{t-1}$ is a random vector with $\mathbb{E}[\bbu^{t-1}] = (\bbp(\Bphi^{t-1}) - \Bphi^{t-1})$.
For comparison, note that we can rewrite \eqref{deterministic update rule homogeneous} as $\Bphi^t = \Bphi^{t-1} + (\bbp(\Bphi^{t-1}) - \Bphi^{t-1})$.
This enables us to apply stochastic approximation~\cite{benaim1999dynamics,Borkar2008} to establish the convergence of the stochastic dynamic.
We summarize our main result for homogeneous market in the theorem below, and leave the discussions of RMA and stochastic approximation to \Cref{sect:analysis},
and the full proof to \Cref{app:RMA}.

\boxthm{[\cite{maldonado2018popularity}, Theorem 5.3]\label{thm-homo-convergence}
In any homogeneous T-O market, if $\Bphi^0 >0$ and $0\leq r < 1$, then with probability $1$,
\begin{equation}
    \lim_{t\rightarrow \infty} \Bphi^t = \Bphi^*,
\end{equation}
where $\Bphi^*$ is the unique interior TOME of the market. When $r\in [0,1]$, with probability $1$,
\begin{equation}
    \lim_{t\rightarrow \infty} \Psi(\Bphi^t) = \Psi^*,
\end{equation}
where $\Psi$ is the objective function of the optimization problem \eqref{Efficiency Entropy Problem},
and $\Psi^*$ is the maximum value of \eqref{Efficiency Entropy Problem}.}

\subsection{TOME for heterogeneous markets}\label{ssec:heteroRMA}

For the heterogeneous case, we first show the following proposition, which depicts that the TOME is optimizing an ex-post version of a convex objective.

\boxprop{\label{prop:Nash-SW}
Given a heterogeneous T-O market with $0 < r_i < 1$ for all $i\in \calU$,
its TOME $\Bphi^*$ is the optimal solution of the following optimization problem:
\begin{equation}
\begin{aligned}
& \max\quad \prod_{i=1}^{|\calU|}\left(\sum_{j=1}^{|\calI|} \q_{ij}\v_{ij}\s_j^{r_i}\right)^{w_ia_i^*},\\
&\text{\emph{subject to} } \Bphi\in\Delta.
\end{aligned}\label{eq:objhetero}
\end{equation}
where $a_i^* = \frac{1}{r_i} \left(\sum_{k=1}^{|\calI|} q_{ik}v_{ik}(\phi^*_k)^{r_i}\right) \left/ \left(\sum_{k=1}^{|\calI|} v_{ik}(\phi^*_k)^{r_i}\right) \right.$ for every $i\in\calU$.}
 
We present the  proof of the above proposition in \Cref{app:hetero}.
The objective function of \eqref{eq:objhetero} takes the form of a product-of-utilities, or sum-of-log-utilities after taking logarithm.
Known as Nash social welfare~\cite{kaneko1979nash}, this objective was found to strike a good balance between fairness and efficiency
in the resulting allocations~\cite{bertsimas2011price,CaragiannisKMPS19}. A proper exposition of this connection is outside the scope of this paper.
Once $a_i^*$ are known (hence \emph{ex-post}), \eqref{eq:objhetero} is a convex optimization problem.
However, if $a_i^*$ is unknown, it is non-convex in general.
We raise the properties of its optimal solution (e.g., is the optimal solution a TOME?) as an open problem.

Then we turn our focus to two interesting special cases of the heterogeneous market, where $r_i = r$ for all types $i$, and:
\begin{itemize}[leftmargin=*]
\item the trial randomness is the same across all user types (i.e., $v_{ij}$ are the same for all types $i$),
but the purchase randomness can be different (i.e., $q_{ij}$ can be different for various types $i$);
\item the trial randomness can be different across all types (i.e., $v_{ij}$ can be different for various types $i$), but the purchase randomness is the same across all types (i.e., $q_{ij}$ are the same for all types $i$).
\end{itemize}

It is easy to reduce the first case to a homogeneous setting.
For the second case, we design a new optimization problem and show that \eqref{cont dynamic} is indeed mirror descent for the problem.
The driving variables of the dynamic \eqref{cont dynamic} are $b_{ij}$ for user types $i$ and items $j$.
We let $b_j := \sum_{i=1}^{|\calU|} b_{ij}$, and set $q_j$ to be the common value of $q_{ij}$ for all $i$.
The optimization problem is

\begin{equation}
\begin{aligned}
&\max \quad \Gamma(\mathbf{b}) := r \sum_{j=1}^{|\calI|} \frac{b_j}{q_j}\log b_j - \sum_{i=1}^{|\calU|} \sum_{j=1}^{|\calI|} \frac{b_{ij}}{q_j}\log \frac{b_{ij}}{q_j v_{ij}}\\
&\text{subject to } \sum_{j=1}^{|\calI|} \frac{b_{ij}}{q_{j}} ~=~ w_i,~\forall i\in \calU,~\text{and}~b_{ij} \ge 0. \label{hetero obj function}
\end{aligned}
\end{equation}

By performing a variable transformation $x_{ij} = b_{ij} / q_j$ to the above problem,
we obtain an equivalent transformed optimization problem where $x_{ij}$'s are the driving variables,
which is needed for the key lemma below. The proof is presented in \Cref{app:hetero}.

\boxlem{\label{Continuous_dynamics_as_mirror_descent}
The dynamic \eqref{cont dynamic} is equivalent to mirror descent w.r.t.~KL divergence on the transformed optimization problem \eqref{hetero obj function}.}

Recall from \Cref{lem:corr-cont-stochastic} that any fixed point of \eqref{cont dynamic} corresponds to a TOME.
\Cref{Continuous_dynamics_as_mirror_descent} thus implies that if the dynamic \eqref{cont dynamic} converges to the optimal solution of \eqref{hetero obj function},
the optimal solution corresponds to a TOME.
Then we use RMA and stochastic approximation again to establish convergence of the corresponding stochastic influence dynamics in heterogeneous markets.
The proof is presented in \Cref{app:RMA}.

\boxthm{\label{thm:convergence-hetero}
In any heterogeneous T-O market with $r_i =r < 1$ for all $i\in \calU$, if $\Bphi^0 >0$, and one of the following conditions
\begin{itemize}
    \item[1.] $v_{ij} = v_j$ for all $i\in \calU, j \in \calI$
    \item[2.] $q_{ij} = q_j$ for all $i\in \calU, j \in \calI$
\end{itemize}
is satisfied, then with probability $1$,
\begin{equation}
    \lim_{t\rightarrow \infty} \Bphi^t = \Bphi^*,
\end{equation}
where $\Bphi^*$ is the unique interior TOME of the market.
When $r\in [0,1]$, with probability $1$,
\begin{equation}
    \lim_{t\rightarrow \infty}\Gamma(\mathbf{b}^t) = \Gamma^*,
\end{equation}
where $\mathbf{b}^t = (b_{ij}^t)$ is the vector with $b_{ij}^t$ defined on $\Bphi^{t-1}$ as in \eqref{cont dynamic},
$\Gamma$ is the objective function of the optimization problem \eqref{hetero obj function},
and $\Gamma^*$ is the maximum value of \eqref{hetero obj function}.
}

\paragraph{A Remark.} In the settings of \Cref{thm-homo-convergence} and \Cref{thm:convergence-hetero},
there indeed exist multiple TOMEs in the simplex $\Delta$. However, there is only one \emph{interior} TOME.
The uniqueness of the limit point of the dynamic depends on the choice of initial point and the social influence parameter $r$:
\begin{itemize}[leftmargin=*]
\item When $r < 1$ and the initial market share $\Bphi^0$ is in the relative interior of $\Delta$,
then the limit point of the dynamic is unique, which is the interior TOME, for both homogeneous and heterogeneous cases
according to Theorems \ref{thm-homo-convergence} and \ref{thm:convergence-hetero}.
\item When $\Bphi^0$ contains zero initial market shares for some items, then it is equivalent to consider a market with those items eliminated.
In other words, those items would not gain non-zero market share in subsequent iterations from zero initial market shares.
The limit point of the dynamic is still unique in these cases.
\item When $r=1$, the objective functions in optimization problems \eqref{Efficiency Entropy Problem} and \eqref{eq:objhetero} are not strictly convex.
There is a level set containing multiple equilibria. The dynamic will converge to the level set which optimizes these objective functions.
\end{itemize}
\section{Analysis}\label{sect:analysis}

We follow the same approach in formally proving two of our main results, \Cref{thm-homo-convergence} and \Cref{thm:convergence-hetero}.
The approach comprises two main steps. First, we show that the evolution of market share $\Bphi$ can be cast as a stochastic RMA.
Second, we show the convergence of such RMA by establishing their equivalence to mirror descent on convex functions.

\subsection{Influence Dynamic as RMA}

\boxdef{(\cite{benaim1999dynamics,robbins1951stochastic})
A Robbins-Monro algorithm (RMA) is a discrete-time stochastic process $z^t$ whose general structure is specified by
\[
\mathbf{z}^{t} - \mathbf{z}^{t-1} ~=~ \gamma^{t} \cdot \left( F(\mathbf{z}^{t-1}) + U^t \right),
\]
where $\mathbf{z}^t\in \mathbb{R}^n$ for some $n\ge 1$, $F:\mathbb{R}^n\rightarrow \mathbb{R}^n$ is a deterministic continuous vector field, $\gamma^t$ is deterministic and satisfies $\gamma^t > 0$, $\sum_{t\ge 1} \gamma^t = +\infty$ and $\lim_{t\rightarrow \infty} \gamma^t = 0$, and $\mathbb{E}[U^t | \mathcal{F}^{t-1}] = 0$ where 
$\mathcal{F}^{t-1}$ is the natural filtration on the entire process.
The corresponding ordinary differential equation (ODE) system of the RMA is $\dot{\mathbf{z}} = F(\mathbf{z})$.}

Note that market share $\Bphi$ will change only when there is a purchase. Thus, Maldonado et al.~\cite{maldonado2018popularity} modify the time schedule to only count those times at which a purchase occurs,
and show the following lemma.

\boxlem{
In the stochastic T-O market, the update of market share follows the following RMA w.r.t.~the modified time schedule:
\[
\Bphi^t - \Bphi^{t-1} ~=~ \frac{1}{t} \cdot \left( (\p(\Bphi^{t-1}) - \Bphi^{t-1}) + U^t \right),
\]
where $U^t$ is the random variable defined as below. Let $\mathbf{e}^t$ denote the random unit vector whose 
$j$-th entry is $1$ if item $j$ is purchased at time $t$. Then $U^t = \mathbf{e}^t - \mathbb{E}[\mathbf{e}^t | \mathcal{F}^{k-1}]$.
(Recall that $e^t_j = 1$ with probability $p_j(\Bphi^{t-1})$.)}

The proof of this lemma can be found in \cite{maldonado2018popularity} (for the homogeneous setting) and \Cref{app:RMA} (for the heterogeneous setting).
With the lemma in hand, we can apply the seminal results of Bena{\"\i}m~\cite{benaim1999dynamics}
to show that the RMA trajectory is the \emph{asymptotic pseudotrajectory} of the mirror descent update \eqref{deterministic update rule homogeneous}.
By using the mirror descent convergence theorem established in \cite{CT93,cheung2018dynamics}, we show that both dynamics converge to the global minimisers of \eqref{Efficiency Entropy Problem}. This allows us to present a new proof of \Cref{thm-homo-convergence}, which was first shown in~\cite{maldonado2018popularity}.

\subsection{Convergence of Mirror Descent}\label{subsect:mirror-descent}

Let $d_h(\cdot,\cdot)$ denote the Bregman divergence in \Cref{defn::Bregmann divergence}.
We assume that the function $h$ is strictly convex. Consequently, $d_h(\cdot,\cdot)$ is strictly convex in its first parameter, and $d_h(\x,\y) = 0$ if and only if $\x = \y$.

\boxdef{A function $f$ is $L$-Bregman-convex with respect to the Bregman divergence $d_h$ if
for any $\y \in \text{rint}(C)$ and $\x\in C$,
\[
f(\y) + \langle\nabla f(\y), \x-\y\rangle ~\leq~ f(\x)  ~\leq~  f(\y) + \langle\nabla f(\y), \x-\y\rangle + L \cdot d_h(\x,\y).\vuppp\vuppp
\]
}

Given an $L$-Bregman-convex function $f$ with respect to the Bregman divergence $d_h$,
the mirror descent rule with respect to the Bregman divergence $d_h$ is given by $\x^{t+1} = g(\x^t)$, where
\begin{equation}
g (\x^t) = \argmin_{\x\in C} \left\{f(\x^{t-1}) + \langle\nabla f(\x^{t-1}), \x - \x^{t-1}\rangle +L \cdot d_h(\x,\x^{t-1}) \right\}. \label{MDMinimizer}
\end{equation}

The update in \eqref{MDMinimizer} is the same as that of a general mirror descent \eqref{general MD rule}, with step-size $\alpha = 1/L$.
It enables us to use the following theorem to bound the difference to the optimal.
\boxthm{[Chen and Teboulle~\cite{CT93}] \label{thm:mirror-descent-1overT}
Suppose $f$ is an $L$-Bregman-convex function with respect to $d_h$, and $\x^t$ is the point reached after $t$ applications of the mirror descent update rule $\x^{t+1} = g(\x^t)$, where $g$ is as defined in \eqref{MDMinimizer}. Then for all $t\ge 1$,
\[
f(\x^t) - f(\x^*) \leq \frac{L\cdot d_h(\x^*, \x^0 )}{t}.\vuppp\vuppp
\]
}

Thankfully, the objective functions for both homogeneous \eqref{Efficiency Entropy Problem} and heterogeneous \eqref{hetero obj function} cases
are Bregman convex; the proofs are presented in \Cref{app:homo,app:hetero} respectively.

Finally, we use the theorem below to complete the proof.
Note that what we have just showed is the convergence of \emph{discrete-time} mirror descent updates of the form $\x^t = g(\x^{t-1})$,
but the theorem requires condition that guarantee convergence of the \emph{continuous-time} ODE system $\dot{\x} = g(\x) - \x$. 
To apply the theorem, we need to convert the discrete-time convergence to its ODE analogue.
The conversion is simple, and it is presented in \Cref{app:RMA}.

\boxthm{Consider an ODE $\dot{\x} = h(\x)$.
Suppose there is a continuously differentiable function $f:\mathbb{R}^d \rightarrow \mathbb{R}$ such that
(i) $\lim_{\|\x\|\rightarrow \infty} f(\x) = +\infty$;
(ii) the set of minimum points of $f$, $X^*$, is non-empty; and
(iii) $\innerProd{\nabla f(\x)}{h(\x)} \le 0$ for all $\x$, with equality holds if and only if $\x\in X^*$.
Then almost surely, the Robbins-Monro algorithm of the ODE converges to a non-empty subset of $X^*$.}
\section{Empirical observations}\label{sect:experiments}

We simulate cultural markets using real-world preferences from the well-known MovieLens dataset~\cite{harper2015movielens}, in order to explore the efficiency and diversity of the market in homogeneous and  heterogeneous settings, and under different ranking strategies\footnote{Code and data that reproduce results in this section is at \url{https://github.com/haiqingzhu543/Stability-and-Efficiency-of-Personalised-Cultural-Markets}}.
he simulations aim to answer key questions such as whether heterogeneous T-O market is more efficient, whether T-O market is stable as prescribed in \Cref{sec:results}, and whether stability sacrifices diversity.

\begin{figure*}[bt]
\centering
\begin{minipage}{0.48\textwidth}
\centering
\includegraphics[width=1.\textwidth]{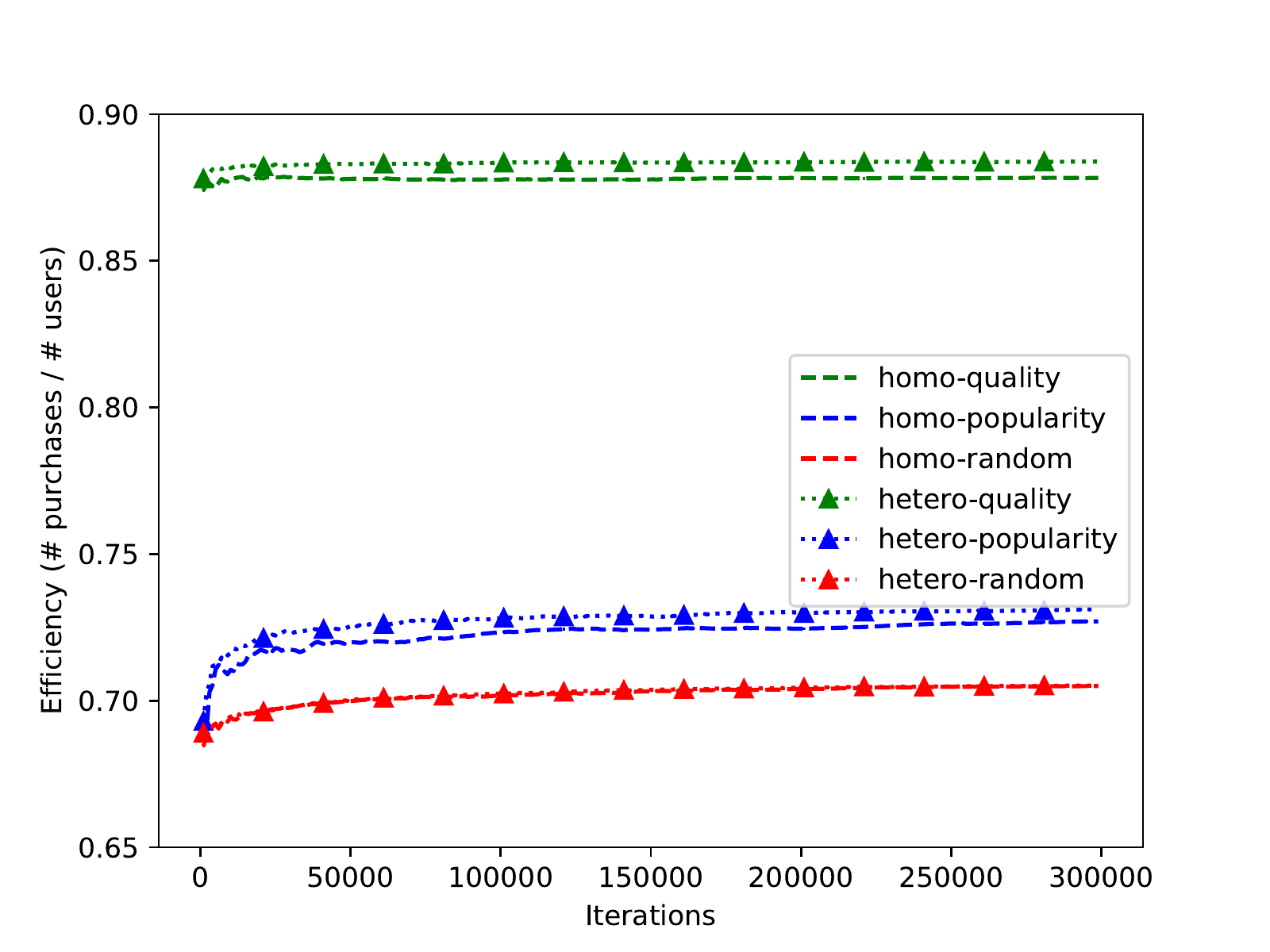}
\end{minipage}
\begin{minipage}{0.48\textwidth}
\centering
\includegraphics[width=1.\textwidth]{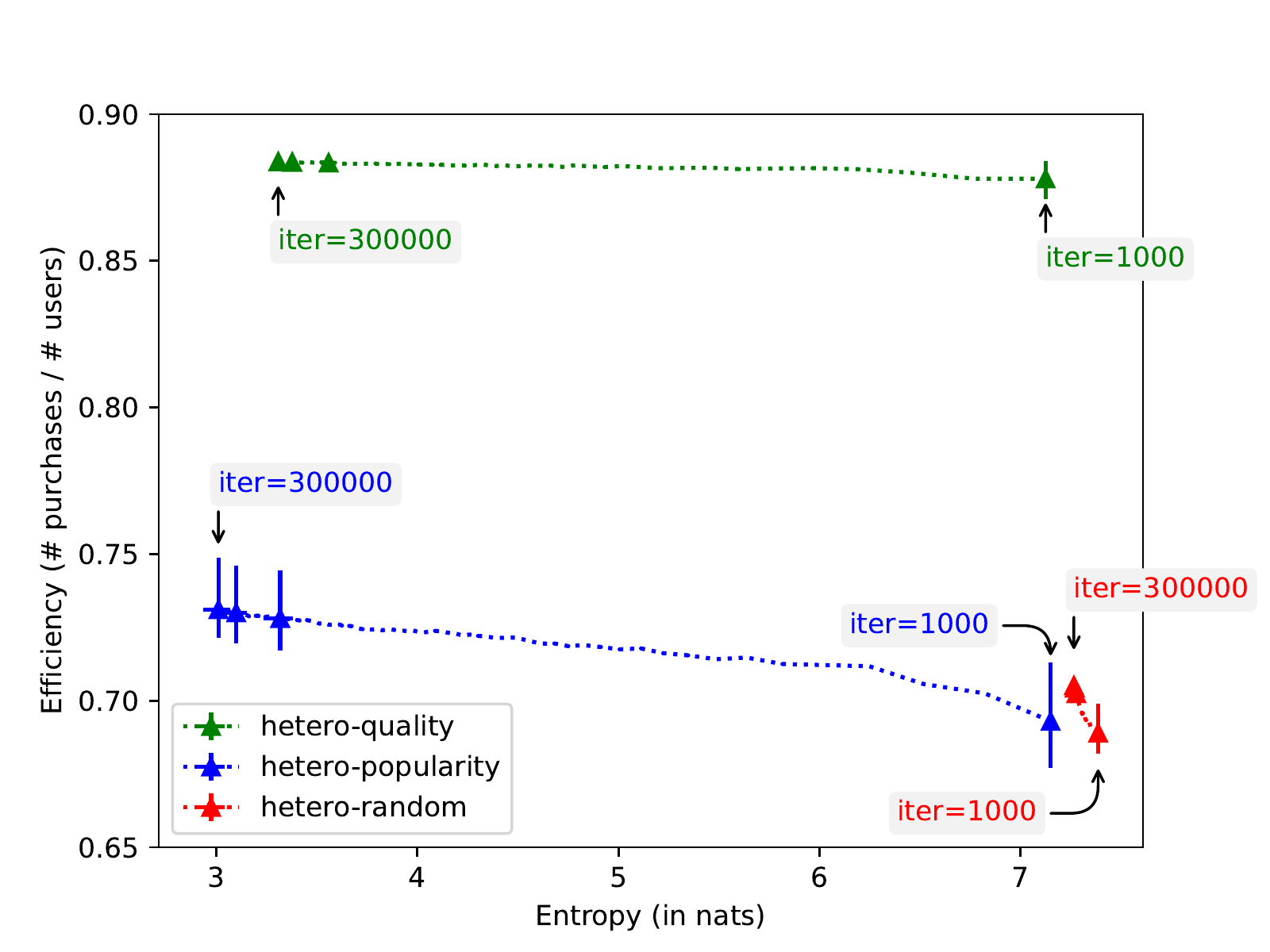}
\end{minipage}

\caption{Simulation results on MovieLens dataset.
Each simulation is run for 300,000 time steps (each introducing a new user), a measurement is taken after every 1000 time steps. (Left) Market efficiency over time comparing the homogeneous vs heterogeneous settings under three ranking strategies (random/popularity/quality). The lines denote the median of 50 simulations with different random initialisations. (Right) Efficiency over the entropy of market shares in heterogeneous setting. The lines denote the median of efficiency over 50 simulations with different random initialisations, markers denote iteration 1000, 100,000, 200,000 and 300,000, and error bars represent the 25th to 75th percentile range in both efficiency and entropy.}
\label{fig:empiricalobs}
\end{figure*}

We set up the simulation using the MovieLens-100K dataset \cite{harper2015movielens}.
This dataset consists of 100,000 ratings (valued 1-5) from 943 users on 1682 movies, where each user has rated at least 20 movies. 
We performed matrix completion using incomplete SVD
\cite{bell2007lessons} via the {\tt Surprise} python package\footnote{\url{https://surpriselib.com}}, yielding a preference matrix $\Gamma = [\gamma_{ij}] \in \mathbb{R}^{943\times1682}$ for each (user, movie) pair. 
With $\gamma_{ij} \in [0,1]$ denoting the normalised preference of  user $i\in \mathcal{U}$ and item (movie) $j \in \mathcal{I}$. 
Denote $\mathcal{O}$ as the set consisting of all indices $(i,j)$ with $\gamma_{i,j}$ observed (in the MovieLens-100K dataset), and  $\overline{\mathcal{O}}$ as the set of unobserved indices (entries estimated with incomplete SVD).

To simulate heterogeneous preference types, we divide the users into $M$ non-overlapping subgroups based on user attributes.
Let $\bigcup_{i=1}^{M}\mathcal{U}_i= \mathcal{U}$, where $\mathcal{U}_i$ denote the set of users in group $i$, and hence the weights $w_i = |\mathcal{U}_i|/ | \mathcal{U}|$. We first calculate the {\em visibility} factor $\hat v_{ij}$ by averaging over the set of observed entries generated by user group $i$ using \eqref{eq:hatVij}. 
We also calculate the {\em quality} factor $q_{ij}$ by averaging over the set of {\it unobserved} entries generated by user group $i$ using \eqref{eq:Qij}.
This choice reflects the intuition that in a T-O market, the probability of purchase depends on a quality factor that is often unknown to the platform a priori before a user tries an item. Other strategies for estimating $q_{ij}$ and $\hat v_{ij}$ are left for future work. 
\begin{eqnarray}
\hat{v}_{ij} =  \frac{1}{| (k,j)\in \mathcal{O}, k \in \mathcal{U}_i |} \sum_{(k,j)\in \mathcal{O}, k \in \mathcal{U}_i} \gamma_{kj}. \label{eq:hatVij}\\
q_{ij} = \frac{1}{|(k,j) \in \overline{\mathcal{O}}, k \in \mathcal{U}_i| }\sum_{(k,j) \in \overline{\mathcal{O}}, k \in \mathcal{U}_i} \gamma_{kj} \label{eq:Qij}
\end{eqnarray}

We cluster users into 100 groups using K-means on the rows of  $\Gamma$.
This grouping is used to compare market efficiency and diversity in homogeneous (all users have the same preferences) and heterogeneous (100 user groups)  settings.
Results for other group compositions are in the online supplement~\cite{supp3065}\ref{app:simulations} and are qualitatively similar.
We also account for position bias in ranked lists~\cite{craswell2008experimental} in order to construct the actual visibility factor $v_{ij}$, which can vary over time, denoted as $v_{ij}^t$. 
Prior {\sc MusicLab} model has included position bias parameters ~\cite{krumme2012QuantifyingSI,maldonado2018popularity} 
for the top-50 items, with zero visibility assigned to all other items.
This results in a list of fixed weights $\iota_k$, $k=1,\ldots,1682$, where $\iota_{1},\ldots,\iota_{50}>0$, and $\iota_{51},\ldots,\iota_{1682} = 0$. 
We adopt the separable click-through rate (CTR) model commonly used in modelling auctions \cite{aggarwal2006truthful}, which simply multiplies the estimated visibility term $\hat{v}_{ij}$ with a ranking factor $\eta_{ij}^t$ of item $j$ presented to user $i$ at time $t$. Different ranking strategies produce different $\eta_{ij}^t$ to modulate $\hat{v}_{ij}$, which in turn result in different probability distributions on the trial phase (equation \eqref{eq:mnlchoice}).

\begin{equation}
v_{ij}^t = \hat{v}_{ij}\eta_{ij}^t
\end{equation}
We introduce the following three ranking strategies that define the relationship between the fixed visibility factors $\iota$ and ranking factor for user $i$ and time $t$ over all items $\eta^t_{i,:}$. In all strategies $\eta_{ij}^t = \iota_k$, with index $k$ defined randomly, ordered by popularity or quality, respectively.

\begin{itemize}[leftmargin=*]
    \item Random-ranking. Upon each simulation round, the visibility term $v_{ij} = \hat v_{ij} \iota_k$, with
    \[ k = \text{random\_permutation} \{ [1,\ldots,|\mathcal{I}|] \} [j]~, \]
    and $[j]$ denotes array indexing. This ranking changes for each simulation step. One expects such a strategy to promote diversity while preserving some information on item appeal through $\hat v_{ij}$.
    \item Popularity-ranking. This strategy sorts the items by descending market share $\Bphi^t$, with
    \[k = \arg \text{sort}_{desc} \{\Bphi^t\} [j]~.\]
    This ranking will change over the simulation steps, and is analogous to the original {\sc MusicLab} experimental setting~\cite{salganik2006ExperimentalSO}. One expects this strategy to be unstable due to the randomness early in the simulation, since it could accidentally promote items that users do not like to the top, resulting in the high quality items being buried. 
    \item Quality-ranking. Denote the descending sorting rank of item $j$ among user group $i$ as $k = \arg \text{sort}_{desc} \{q_{i, :}\} [j]$, where $q_{i, :}$ is the one-dimensional array for qualities factors in group $i$. 
    This ranking does not change over the simulation steps, since both $q_{ij}$ and its sorted order remains fixed. One expects this strategy to best align visibility with the underlying quality metrics (unobserved before trying), since it has {\em oracle} access to $q_{ij}$, and should yield high efficiency. 
\end{itemize}

In each round of the simulation, one new user arrives at the market and chooses an item for a \emph{trial} according to the multinomial logit (equation \eqref{eq:mnlchoice}). Then the user decides whether to \emph{purchase} this particular item by flipping a biased coin parameterized by $q_{ij}$. 
Note that these new users are {\it generalisations} of the $M$ groups of user populations in MovieLens via attributes $q_{ij}$ and $\hat v_{ij}$, rather than being subsets (or samples) from the original 943 users.
This setting is consistent with other theoretical and simulation studies of cultural market and recommender systems~\cite{krumme2012QuantifyingSI,jiang2019degenerate}. We report {\em market efficiency} as the empirical version of \Cref{defn:efficiency}, namely, the fraction of users who made a purchase. We also measure diversity among the set of items, by computing the Shannon entropy of market share $\Bphi$ (\Cref{eq:entropy}) at each time step. We explore the relationship between these two metrics.

\Cref{fig:empiricalobs} summarises the trends of efficiency and diversity over the two settings -- homogeneous and heterogeneous -- and three ranking strategies -- random/popularity/quality. In \Cref{fig:empiricalobs} (left), it is observed that the quality-ranking oracle has the highest efficiency among the three ranking strategies, followed by popularity ranking, and random ranking has the lowest efficiency when there is a sufficient number of users. 
Taking into account heterogeneous user preferences improves efficiency in both quality and popularity ranking settings. We also notice that in MovieLens-100K dataset, the gap among the ranking strategies are larger than that of moving from the homogeneous to heterogeneous settings.
\Cref{fig:empiricalobs} (right) compares both efficiency and diversity (measured by the entropy of market shares) at different iterations for different ranking strategies in the heterogeneous setting. 
For popularity and quality rankings, item diversity decreases over time (curves moving leftwards) while efficiency increases (curves moving slightly up). 
Comparing the two, quality ranking yields more diversity across the items (higher entropy) and is more stable (smaller spread on both dimensions). 
This observation corroborates \Cref{prop:Nash-SW} that heterogeneous Musiclab objective is ex-post concave. 
Popularity ranking results in larger variations in both efficiency and entropy, confirming observations in the original \textsc{MusicLab} experiment~\cite{salganik2006ExperimentalSO} -- that market allocation is unstable due to random initialisations and result in market dominance by a few popular items.
As a control group, random ranking yields the lowest efficiency and no apparent differences between homogeneous and heterogeneous user preferences. In this setting, efficiency still improves slightly due to the joint effect of both visibility and quality terms. But item diversity stays close to the theoretical maximum in entropy ($\ln (1684) \sim 7.4 nats$) over time, indicating that the random ranking with cut-off at top 50 items is playing a larger role in user choice than signals present in the visibility and quality terms.
\section{Conclusion}\label{sec:conclusion}

This paper views the dynamics of cultural markets under an optimization lens. 
We identify new objective functions for trial-offer markets, and establish robust connections between social feedback signals and optimization processes.
Our results narrow the gap between the theory and practice of recommender systems. In particular, they make the analysis of recommender systems more versatile by incorporating user-specific preferences, and offer a holistic view of market stability and efficiency  beyond individual clicks and views. 
Simulations using real-world user preferences confirm that markets with heterogeneous preferences are more stable and more efficient.

Our work leads to several open research questions, such as convergence rates of the stochastic T-O markets, analysis of general heterogeneous T-O settings, fairness properties of market equilibria, and describing markets that are also learning a recommender systems in-the-loop \cite{mladenov2021recsim}. More generally, we hope the current work opens up new ways to asking and answering a set of research questions at the intersection of classical markets and online attention.

\section*{Acknowledgment}
LX and HZ are supported in part by AFOSR project FA2386-20-1-4064. Views expressed in this article are those of the authors and not of the funding agency.
We thank Alvaro Flores for discussions on choice models and \textsc{MusicLab}. 

\bibliographystyle{plain}
\bibliography{refs}

\newpage
\appendix


\section{Properties of TOME}\label{app:property}

In this section, we present the proofs of \Cref{thm:exist-TOME} and \Cref{lem:corr-cont-stochastic}.

\subsection{Proof of \Cref{thm:exist-TOME}}

By the definition of $y_j(\Bphi)$ in \eqref{choice TO hetero},
the following inequality holds for any $\Bphi \in \Delta$ and $j\in \calI$:
\[
y_j(\Bphi) ~=~ \sum_{i=1}^{|\calU|} w_i q_{ij} \cdot \frac{v_{ij} (\phi_j)^{r_i}}{\sum_{k=1}^{|\calI|} v_{ik} (\phi_k)^{r_i}} ~\ge~ \underbrace{\frac{w_{i^*(j)} v_{i^*(j),j} q_{i^*(j),j}}{\sum_{k=1}^{|\calI|} v_{i^*(j),k}}}_{c_j}\cdot (\phi_j)^{r_{i^*(j)}}~,
\]
where $i^*(j)$ is a user type with $v_{i^*(j),j} q_{i^*(j),j} > 0$, which exists due to condition (ii).
If $\phi_j \ge (c_j)^{1/(1-r_{i^*(j)})}$, then 
\begin{equation}\label{y fixed point}
y_j(\Bphi) \ge c_j \cdot (c_j)^{r_{i^*(j)}/(1-r_{i^*(j)})} = (c_j)^{1/(1-r_{i^*(j)})}.
\end{equation}

On the other hand, recall that $p_j(\Bphi) = y_j(\Bphi) / (\sum_{k=1}^{|\calI|} y_k(\Bphi))$.
Since $\sum_{k=1}^{|\calI|} y_k(\Bphi) \le 1$, 
\begin{equation}\label{p fixed point}
p_j(\Bphi) \ge y_j(\Bphi).
\end{equation}

Let $S$ denote the set $\left\{\Bphi\in \Delta : \forall j\in \calI,~1\ge \phi_j\ge (c_j)^{1/(1-r_{i^*(j)})}\right\}$.
Since $0 < c_j \le 1$, $1/(1-r_{i^*(j)}) \ge 1$ and $0\le q_{ij}\le 1$,
\[
\sum_{j=1}^{|\calI|}(c_j)^{1/(1-r_{i^*(j)})} ~\le~ \sum_{j=1}^{|\calI|} c_j ~=~ \sum_{i=1}^{|\calU|} \sum_{j:i^*(j)=i} c_j ~\le~ \sum_{i=1}^{|\calU|} w_i ~=~ 1,
\]
hence $S$ is non-empty. And of course, $S$ is compact and convex.
Due to \eqref{y fixed point} and \eqref{p fixed point}, $\p$ is a continuous function that maps $S$ to a subset of $S$.
By the Brouwer's fixed point theorem, $\p$ has a fixed point in $S$, which is a TOME of the market.

\subsection{Proof of \Cref{lem:corr-cont-stochastic}}

\noindent (i) Suppose $\Bphi^* = (\phi_j^*)$ is a TOME. By the definition of TOME, there exists a real number $c$ such that
for any $j\in \calI$,
\begin{equation}\label{eq:Phis-ys}
\phi^*_j = c\cdot y_j(\Bphi^*).
\end{equation}

Recall that we set $b_{ij}^* = w_i q_{ij} \cdot \frac{v_{ij} (\phi^*_j)^{r_i}}{\sum_{k=1}^{|\calI|} v_{ik} (\phi^*_k)^{r_i}}$.
Let $b_j^* := \sum_{i=1}^{|\calU|} b_{ij}^*$, then by \eqref{choice TO hetero},
\begin{equation}\label{eq:bs-ys}
b_j^* = y_j(\Bphi^*).
\end{equation}

Now, suppose that $\bbb^{t-1} = \mathbf{b}^*$. By \eqref{cont dynamic}, \eqref{eq:Phis-ys} and \eqref{eq:bs-ys},
\[
b_{ij}^t ~=~ w_i q_{ij} \cdot \frac{v_{ij} (b^*_j)^{r_i}}{\sum_{k=1}^{|\calI|} v_{ik} (b^*_k)^{r_i}}
~=~ w_i q_{ij} \cdot \frac{v_{ij} (\phi_j^*)^{r_i}}{\sum_{k=1}^{|\calI|} v_{ik} (\phi_k^*)^{r_i}} ~=~ b_{ij}^*.
\]
Thus, if $\bbb^{t-1} = \bbb^*$, then $\bbb^t = \bbb^*$ too.
This concludes that $\bbb^*$ is a fixed point of the dynamic \eqref{cont dynamic}.

\medskip

\noindent (ii) Note that
\[
\begin{aligned}
b_j^* ~:=~ \sum_{i=1}^{|\calU|} b_{ij}^* &~=~ \sum_{i=1}^{|\calU|} w_i q_{ij} \cdot \frac{v_{ij} (b^*_j)^{r_i}}{\sum_{k=1}^{|\calI|} v_{ik} (b^*_k)^{r_i}} ~~~~\text{(since $\bbb^*$ is a fixed point of \eqref{cont dynamic})}\\
&~=~ \sum_{i=1}^{|\calU|} w_i q_{ij} \cdot \frac{v_{ij} (\phi^*_j)^{r_i}}{\sum_{k=1}^{|\calI|} v_{ik} (\phi^*_k)^{r_i}} ~~~~\text{(since $\phi_j^* := b_j^* /(\sum_{k=1}^{|\calI|} b_k^*)$ for all $j$)}\\
&~=~ y_j(\Bphi^*).
\end{aligned}
\]
Thus, for any item $j$ we have 
$\phi_j^* ~:=~ \frac{b_j^*}{\sum_{k=1}^{|\calI|} b_k^*} ~=~ \frac{y_j(\Bphi^*)}{\sum_{k=1}^{|\calI|} y_k(\Bphi^*)}$,
which implies $\Bphi^*$ is a TOME by definition.

\section{Homogeneous Markets}\label{app:homo}

\subsection{Proof of Theorem \ref{thm:social-welfare}}

\subsubsection*{Optimization Problem \eqref{TotalUtilityProblem}}

When $r \ge 1$, since $\phi_j \in [0,1]$, we have $\phi_j^r \leq \phi_j$. Let $\hat{q} = \max_j\{\bar{q}_j: j\in \calI\}$. We have
\[
\sum_j^{|\calI|} \bar{q}_j\phi_j^r ~\leq~ \sum_j^{|\calI|} \bar{q}_j\phi_j ~\leq~ \hat{q} \sum_j^{|\calI|} \phi_j ~=~ \hat{q},
\]
If $r > 1$, the above equalities hold only if $\phi_j = 1$ for an item $j$ satisfying $q_j = \hat{q}$, and $\phi_k = 0$ for all $k\neq j$.
It is easy to verify that every such $\Bphi$ is a TOME (check \ref{defn:TOME}).
If $r=1$, the above equalities hold only if: $\phi_j > 0 ~\Rightarrow~ q_j = \hat{q}$. Again, it is easy to verify that every such $\Bphi$ is a TOME.

When $0 < r < 1$, \eqref{TotalUtilityProblem} is a convex program, so we can employ the standard convex analysis tools of Lagrangian multipliers and Karush–Kuhn–Tucker (KKT) theorem to characterize the optimal solution.
We first transform the maximisation problem into a minimisation problem for simplicity:
\[
\begin{aligned}
&\min \quad -\sum_j^{|\calI|} \bar{q}_j\phi_j^r, \\
&\text{subject to } \Bphi\in\Delta.
\end{aligned}
\]
The Lagrangian is given below, with dual variables $\lambda\in \mathbb{R}$ and $\eta_j\ge 0$ for any $j\in \calI$.
\[
\mathcal{L}_1 =  -\sum_j^{|\calI|} \bar{q}_j\phi_j^{r} + \lambda\left(\sum_{j= 1}^{|\calI|} \phi_j - 1\right) - \sum_{j= 1}^{|\calI|} \eta_j \phi_j.
\]
Then for any $j\in \calI$,
\[
\frac{\partial\mathcal{L}_1  }{\partial \phi_j} ~=~ -r\bar{q}_j\phi_j^{r-1} + \lambda - \eta_j ~=~0
\]
Since $r-1 < 0$, for the above equality to hold, $\phi_j = 0$ is impossible if $\bar{q}_j > 0$.
By the KKT theorem, at the optimum $\eta_j = 0$ for any $j$ with $\bar{q}_j > 0$, and hence $-r\bar{q}_j\phi_j^{r-1} + \lambda = 0$.
Thus, $\frac{\phi_j^{1-r}}{\bar{q}_j}$ is the same for all $j$ with $\bar{q}_j > 0$ (with common value $r-1-\lambda$), i.e., $\phi_j^{1-r}\propto \bar{q}_j$.
On the other hand, if $\bar{q}_j = 0$, then clearly $\phi_j = 0$ at the optimum.
Together with the constraint $\sum_{j= 1}^{|\calI|} \phi_j$, we can solve for the optimal solution analytically, which coincides with the TOME specified in \eqref{eq:TOMEhomo}.

\subsubsection*{Optimization Problem \eqref{Efficiency Entropy Problem}}

Note that the objective function can be decomposed as two parts, $\sum_j \phi_j \log \bar{q}_j$ and $(1-r)$ times the Shannon entropy.
When $r>1$, $1-r < 0$, so the second part is maximized when the market share is purity, i.e., $\phi_j = 1$ for some $j$ and $\phi_k = 0$ for all $k\neq j$.
The first part is maximized when $\phi_j = 1$ for some $j$ satisfying $q_j = \hat{q}$, and $\phi_k = 0$ for all $k\neq j$ -- note that this condition is stronger than the one for the second part,
and this is the same optimality condition as the one we presented for \eqref{TotalUtilityProblem}.

When $r=1$, the optimality condition is again the same as what we presented for \eqref{TotalUtilityProblem}: $\phi_j > 0 ~\Rightarrow~ q_j = \hat{q}$.

When $0<r<1$, the objective function is concave, so \eqref{Efficiency Entropy Problem} is a convex program. The Lagrangian is
\[
\mathcal{L}_2 =  -\sum_j^{|\calI|} \phi_j \log \bar{q}_j + (1-r)\phi_j \log \phi_j + \lambda\left(\sum_{j= 1}^{|\calI|} \phi_j - 1\right) - \sum_{j= 1}^{|\calI|} \eta_j \phi_j.
\]
Then for any $j\in \calI$,
\[
\frac{\partial\mathcal{L}_2}{\partial \phi_j} ~=~ -\log \bar{q}_j + (1-r) (1+\log \phi_j) + \lambda - \eta_j ~=~ 0.
\]
Since $\log \phi_j \searrow -\infty$ as $\phi_j \searrow 0$, $\phi_j = 0$ is impossible.
By the KKT theorem, at the optimum $\eta_j = 0$ for all $j$. Thus,
$\log \frac{\phi_j^{1-r}}{\bar{q}_j}$ is the same for all $j$ (with common value $r-1-\lambda$), i.e., $\phi_j^{1-r}\propto \bar{q}_j$.
This is the same optimality condition as the one we presented for \eqref{TotalUtilityProblem}.

\subsection{Proof of Lemma \ref{lem:mirror-descent-homo}}

Mirror descent is used to minimize a function, so to use the mirror descent optimization tools more conveniently, in the proofs below
we let $\Psi$ be the negative of the objective function in \eqref{Efficiency Entropy Problem}. Note that
\[
\frac{\partial \Psi(\Bphi_j^{t-1})}{\partial \phi_j} ~=~ -\log \bar{q}_j + (1-r) (1+\log \phi_j^{t-1})~.
\]
Thus, the mirror descent update rule is\footnote{We have removed terms that does not depend on $\Bphi$ from the function in the $\argmin$, since they do not affect the update rule.}
\[
\Bphi^t = \argmin_{\Bphi\in \Delta} \left\{ \sum_{j=1}^{|\calI|} \left( -\log \bar{q}_j + (1-r) (1+\log \phi_j^{t-1}) \right)\cdot \phi_j + \left(\phi_j \log \frac{\phi_j}{\phi_j^{t-1}} - \phi_j\right) \right\}
\]
Let the function in the $\argmin$ be $g$. Then
\[
\frac{\partial g}{\partial \phi_j} = \log \frac{\phi_j}{\bar{q}_j (\phi_j^{t-1})^r} + (1-r)~.
\]
To optimize $g$ in the simplex $\Delta$, from KKT condition it suffices that the above partial derivative is the same for all $j$, i.e.,
$\frac{\phi_j}{\bar{q}_j (\phi_j^{t-1})^r}$ is the same for all $j$. Together with the constraint $\Bphi \in \Delta$, we can solve for $\Bphi$ analytically,
which is $\p(\Bphi^{t-1})$ as given by \eqref{choice TO}.

\subsection{Convergence in Deterministic Homogeneous T-O Market}

For fulfilling our final target of showing convergence in stochastic T-O market, an intermediate conceptual step is to show the analogous result in the corresponding deterministic T-O market.
After proving Lemma \ref{lem:mirror-descent-homo} above, we use the approach laid down in \Cref{subsect:mirror-descent} to show the update rule \eqref{deterministic update rule homogeneous}
converges to TOME. To apply \Cref{thm:mirror-descent-1overT}, we need to show that $\Psi$ (again, here it denotes the negative of the objective function of \eqref{Efficiency Entropy Problem}) is $1$-Bregman convex w.r.t.~the KL divergence. For any $\Bphi,\Bphi'\in \Delta$,

\[
\begin{aligned}
\Psi(\Bphi') - \Psi(\Bphi) - \langle\nabla \Psi(\Bphi), \Bphi' - \Bphi\rangle
&=~ \sum_{j=1}^{|\calI|} (\phi_j - \phi'_j) \log \bar{q}_j + (1-r) (\phi'_j \log \phi'_j - \phi_j \log \phi_j) \\
& \hspace*{1in} + \left[ \log \bar{q}_j - (1-r) (1+\log \phi_j)  \right](\phi'_j - \phi_j)\\
&=~ (1-r)\sum_{j=1}^{|\calI|} \phi'_j \log \phi'_j - \phi'_j \log \phi_j\\
&=~ (1-r)\cdot \textsc{KL}(\Bphi',\Bphi)~\le~ \textsc{KL}(\Bphi',\Bphi).
\end{aligned}
\]

\section{Heterogeneous Markets}\label{app:hetero}

\subsection{Proof of Proposition \ref{prop:Nash-SW}}

Let $f$ be the logarithm of the objective function, then the partial derivatives are given by
\[
\frac{\partial f}{\partial \s_j} = \sum_{i=1}^{|\calU|} w_ia_i^*r_i\frac{\q_{ij}\v_{ij}\s_j^{r_i-1}}{\sum_{k=1}^{|\calI|}\q_{ik}\v_{ik}\s_k^{r_i}}.
\]
We note that if $\phi^*_j =0$ for some $j\in \calI$, then this component clearly satisfies the equilibrium equation.
Therefore, it suffices to consider the set $Q \subset \calI$ which includes all item indices $j$ with $\phi^*_j > 0$.
We have $\sum_{j\in Q} \phi^*_j= 1$. By the KKT theorem, we have
 \begin{equation}
 	 \sum_{i=1}^{|\calU|} w_ia_i^*r_i\frac{\q_{ij}\v_{ij}\s_j^{r_i-1}}{\sum_{k=1}^{|\calI|}\q_{ik}\v_{ik}\s_k^{r_i}} + \lambda = 0 \label{KKT_multi}
 \end{equation} 
 for all $j\in Q$. Multiplying \eqref{KKT_multi} by $\phi_j$ and summing it up over all $j\in Q$ gives
\[
\begin{aligned}
0 ~& =~ \sum_{j\in Q}\left(\sum_{i=1}^{|\calU|} w_ia_i^*r_i\frac{\q_{ij}\v_{ij}\phi_j^{r_i}}{\sum_{k=1}^{|\calI|}\q_{ik}\v_{ik}\phi_k^{r_i}} + \lambda\phi_j\right) \\
&=~ \sum_{i=1}^{|\calU|} w_ia_i^*r_i\frac{\sum_{j\in Q}\q_{ij}\v_{ij}\phi_j^{r}}{\sum_{k=1}^{|\calI|}\q_{ik}\v_{ik}\phi_k^{r_i}} ~+~ \lambda \\
&=~ \sum_{i=1}^{|\calU|} w_ia_i^*r_i + \lambda,
\end{aligned}
\]
or $\lambda = -\sum_{i=1}^{|\calU|} w_ia_i^*r_i$. Plugging this into \eqref{KKT_multi} we get
\[
\phi_j ~=~ \sum_{i=1}^{|\calU|} \frac{w_ia_i^*r_i}{\sum_{h=1}^{|\calU|}w_ha_h^*r_h} \cdot \frac{\q_{ij}\v_{ij}\s_j^{r_i}}{\sum_{k=1}^{|\calI|}\q_{ik}\v_{ik}\s_k^{r_i}},
\]
or equivalently
\[
\phi_j ~\propto~ \sum_{i=1}^{|\calU|} w_ia_i^*r_i \cdot \frac{\q_{ij}\v_{ij}\s_j^{r_i}}{\sum_{k=1}^{|\calI|}\q_{ik}\v_{ik}\s_k^{r_i}}.
\]
By setting $a_i^* = \frac{1}{r_i} \cdot \frac{\sum_{k=1}^{|\calI|} q_{ik}v_{ik}(\phi^*_k)^{r_i}}{\sum_{k=1}^{|\calI|} v_{ik}(\phi^*_k)^{r_i}}$,
when setting $\Bphi = \Bphi^*$ in the RHS of the above formula, it fulfills the definition of TOME (recall \Cref{lem:market-eff} and \Cref{defn:TOME}).

\subsection{Reduction to Homogeneous Market for $\v_{ij} = \v_j$}\label{Special case: $\v_{ij} = \v_j$ for heterogeneous setup}
Recall that in a T-O market each user tries an item modelled by a stochastic process, followed by a random decision to purchase that item or not.
In the heterogeneous setting, if for each item $j$ we have $v_{ij} = v_j$ for all $i\in \calU$, this means all users follow the same stochastic process of choosing which item to try.
Then the model is equivalent to all users belong to the same type (i.e., homogeneous), but the eternal probability $\tilde{q}_j$ of purchasing an item $j$ after trying is a weighted average of $q_{ij}$'s over all $i\in \calU$. The following proposition formally summarises the above idea.

\begin{prop}
Suppose that $r_i = r$ for all $i\in \calU$. Also, for every item $j \in \calI$, we have $v_{ij} = v_j$ for all $i\in \calU$.
Then the TOME $\Bphi^*$ can be written as
\[
\phi^*_j = \frac{(\tilde{q}_j\v_j)^{\frac{1}{1-r}}}{\sum_{j=1}^{|\calI|}(\tilde{q}_j\v_j)^{\frac{1}{1-r}}},
\]
where $\tilde{q}_j = \sum_{i=1}^{|\calU|}w_i q_{ij}$.
 \end{prop}

\begin{pf}
When $r_i = i$ and $v_{ij} = v_j$, the function $y_j(\Bphi)$ in \eqref{choice TO hetero} can be written as
\[
\sum_{i=1}^{|\calU|} w_i q_{ij} \cdot \frac{v_j (\phi_j)^r}{\sum_{k=1}^{|\calI|} v_k (\phi_k)^r} ~=~ 1\cdot \left( \sum_{i=1}^{|\calU|} w_i q_{ij} \right)
\cdot \frac{v_j (\phi_j)^r}{\sum_{k=1}^{|\calI|} v_k (\phi_k)^r}~.
\]
The RHS is same as $y_j(\Bphi)$ for a homogeneous market with $w_1 = 1$, the same values of $v_j$'s, and with $q_j = \sum_{i=1}^{|\calU|} w_i q_{ij}$.
\end{pf}

\subsection{Proof of Lemma \ref{Continuous_dynamics_as_mirror_descent}}

First, we write down the transformed optimization problem explicitly. Recall that $\sum_{i=1}^{|\calU|} w_i = 1$, and $x_{ij} = b_{ij} / q_j$. Let $x_j = \sum_{i=1}^{|\calU|} x_{ij}$.
\begin{equation}
\begin{aligned}
&\min \quad \Gamma(\X) = -r \sum_{j=1}^{|\calI|} x_j \log (x_j q_j) + \sum_{i=1}^{|\calU|} \sum_{j=1}^{|\calI|} x_{ij} \log \frac{x_{ij}}{v_{ij}},\\
&\text{subject to } \sum_j x_{ij} = w_i,\text{ and } x_{ij}\geq 0 \text{  for all $i,j$}.
\end{aligned}\label{OBJ-MirrorDesecnt}
\end{equation}
Note that $\sum_{i,j} x_{ij} = \sum_i w_i = 1$ due to the constraints. We first show that the objective function is $1$-Bregman-convex w.r.t.~the KL divergence on the variables $\X$.
Note that
\[
\frac{\partial \Gamma}{\partial x_{ij}} = -r \log (x_j q_j) + \log \frac{x_{ij}}{v_{ij}} + 1 - r.
\]
Then for any $\X,\X'$ in the domain, a direct calculation shows that
\[
\Gamma(\X') - \Gamma(\X) - \langle \nabla \Gamma(\X),\X'-\X\rangle = \textsc{KL}(\X',\X) - r\cdot \textsc{KL}(\mathbf{y}',\mathbf{y})~,
\]
where $\mathbf{y} = (x_1,x_2,\ldots,x_{|\calI|})$ and $\mathbf{y}' = (x'_1,x'_2,\ldots,x'_{|\calI|})$. Since $\X,\X'$ are refinements of $\mathbf{y},\mathbf{y}'$ respectively,
$0 \le \textsc{KL}(\mathbf{y}',\mathbf{y}) \le \textsc{KL}(\X',\X)$. Since $0 < r < 1$,
\[
0~\le~\Gamma(\X') - \Gamma(\X) - \langle \nabla \Gamma(\X),\X'-\X\rangle ~\le~ \textsc{KL}(\X',\X),
\]
which demonstrates that $\Gamma$ is $1$-Bregman-convex w.r.t.~the KL divergence.

Next, we compute the mirror descent update rule derived from \eqref{MDMinimizer}, which is
\[
\X^{t} ~=~ \argmin_{\X} \left\{ \sum_{i,j} \left( -r \log (x_j^{t-1} q_j) + \log \frac{x_{ij}^{t-1}}{v_{ij}} + 1 -r \right) x_{ij} ~+~ \textsc{KL}(\X,\X^{t-1}) \right\}~.
\]
Let the function in the $\argmin$ be $g$. Then
\[
\frac{\partial g}{\partial x_{ij}} ~=~ -r \log (x_j^{t-1} q_j) + \log \frac{x_{ij}^{t-1}}{v_{ij}} + 1 -r + \log \frac{x_{ij}}{x_{ij}^{t-1}}
~=~ \log \frac{x_{ij}}{v_{ij} (q_j x_j^{t-1})^r} + 1-r.
\]
To optimize $g$ subject to the constraints in the optimization problem, from KKT condition it suffices that for each user type $i$,
the above partial derivative is the same for all $j$, i.e., $\frac{x_{ij}}{v_{ij} (q_j x_j^{t-1})^r}$ is the same for all $j$.
Together with the constraint $\sum_j x_{ij} = w_i$, we can derive the optimal solution, which is
\[
x_{ij}^t ~=~ w_i\cdot \frac{v_{ij} (q_j x_{j}^{t-1})^r}{\sum_{k=1}^{|\calI|} v_{ik} (q_k x_{k}^{t-1})^r}~.
\]
Finally, recall that we have adopted the variable substitution $x_{ij} = b_{ij} / q_j$. Converting the above update rule back to the domain with driving variables $b_{ij}$'s, we have
\[
b_{ij}^t ~=~ w_i q_j\cdot \frac{v_{ij} (b_{j}^{t-1})^r}{\sum_{k=1}^{|\calI|} v_{ik} (b_{k}^{t-1})^r}~,
\]
which matches with \eqref{cont dynamic}.

\newpage
\section{The Robbins-Monro Algorithm and Stochastic Approximation}\label{app:RMA}

\subsection{Background}

A dynamical system is typically modelled as a \emph{continuous-time} differential equation system.
While this matches with the reality well for many natural systems (e.g., physics laws),
other systems that involve humans and computers have \emph{discrete-time} updates, and are often subject to noises and randomness.
Stochastic approximation is a subject that studies these discrete-time stochastic systems,
with fascinating results that establish connections to their continuous-time analogs.
We give a brief discussion of the background, which is mainly extracted from Bena{\"\i}m~\cite{benaim1999dynamics} and Borkar~\cite{Borkar2008}.
We start with the definition of Robbins-Monro algorithm.

\boxdef{[Bena{\"\i}m~\cite{benaim1999dynamics}, Section 4.2]
Let $(\Omega,\mathcal{F},\mathbb{P})$ be a probability space and $\{\mathcal{F}_n\}_{n\in \mathbb{N}} \subset \mathcal{F}$ a filtration.
A stochastic process $\{\bbz^t\}_{t\in \nn} \subset \{\rr^m\}^{\nn}$ is a Robbin-Monro algorithm (RMA) if it is in the form of
\begin{equation}\label{eq:robbins-monro-general}
\bbz^t = \bbz^{t-1} + \gamma^t\cdot (F(\bbz^{t-1}) + \bbu^t),
\end{equation}
where $F: \mathbb{R}^m \rightarrow \mathbb{R}^m$ is a continuous function (vector field), and $\{\bbu^t\}_{t\in\nn}$ is adapted:
for all $t\ge 1$, $\bbu^t$ is measurable w.r.t.~$\calF_{t-1}$. Furthermore,
\begin{itemize}
\item[(i)] $\{\gamma^t\}_{t\in\nn}$ is a deterministic sequence;
\item[(ii)] $\mathbb{E}[\bbu^t ~|~\calF_{t-1}] = \mathbf{0}$. \vuppp\vuppp
\end{itemize} 
}

To describe the connections between discrete-time stochastic systems and their continuous-time analogs,
we need the notions of \emph{semiflow} and \emph{asymptotic pseudotrajectory}.

\boxdef{[Bena{\"\i}m~\cite{benaim1999dynamics}, Section 3]
A semiflow $\Phi$ on a metric space $(M,d)$ is a continuous map
\begin{equation}
\begin{aligned}
\Phi: \mathbb{R}_+ \times M \rightarrow M, \\
\Phi(t,\bbz) = \Phi_t(\bbz)
\end{aligned}
\end{equation}
such that
\begin{equation}
\Phi_0 = \textsf{Id}, \ \Phi_{t+s} = \Phi_{t} \circ \Phi_{s}
\end{equation}
for all $(t,s) \in \mathbb{R}_+ \times \mathbb{R}_+$.
}

A prominent example of semiflow is dynamical system of the form $\dot{\bbz} = F(\bbz)$. We call such a system a semiflow induced by the vector field $F$.
Roughly speaking, an asymptotic pseudotrajactory is a trajectory which is very close to a semiflow if we push $t$ to infinity.

\boxdef{[Bena{\"\i}m~\cite{benaim1999dynamics}, Section 3]
A continuous function $Z: \rrplus \rightarrow M$ is an asymptotic pseudotrajectory for a semiflow $\Phi$ if
\begin{equation}
\lim_{t \rightarrow \infty}~\sup_{0\leq h \leq T}~d(Z(t+h) , \Phi_h(Z(t))) = 0
\end{equation}
for any $T>0$. 
}

Suppose a RMA \eqref{eq:robbins-monro-general} in which the step-sizes $\{\gamma^t\}_{t\in\nn}$ satisfy
\begin{equation}
\sum_{n=1}^{\infty}\gamma_n = \infty,\quad\lim_{n\rightarrow \infty} \gamma_n = 0~. \label{RMA-stepsizes}
\end{equation}
Let $\tau_0 = 0$ and for $t\ge 1$, let $\tau_t = \sum_{i=1}^{t}\gamma^i$.
The continuous-time affine interpolated process $Z(t)$ is, for any $t\ge 0$ and $ 0\le s \le \tau_{t}-\tau_{t-1}$,
\begin{equation}
Z(\tau_{t-1}+s) = z^{t-1} +s\cdot \frac{z^t - z^{t-1}}{\tau_t - \tau_{t-1}}.
\end{equation}

The following powerful theorem establishes the connection between RMA and its corresponding semiflow.

\boxthm{[Bena{\"\i}m~\cite{benaim1999dynamics}, Section 4] \label{thm:sa-benaim}
Let $F$ be a smooth vector field on $M$. In addition, for any point $p\in M$ it has unique integral curves around $p$.
If the step-sizes of a RMA satisfies \eqref{RMA-stepsizes}, and
\begin{enumerate}
\item for some $q\ge 2$, $\sup_n \mathbb{E}[\Vert \bbu^t\Vert^q] < \infty$ and $\sum_{n\in\mathbb{N}} \gamma_n^{1+\frac{q}{2}} < \infty$;
\item either $\sup_n{\Vert x_n\Vert} < \infty$, or $F$ is Lipschitz and bounded on a neighbourhood of $\{x_n: n\geq 0\}$,
\end{enumerate}
then the interpolated process of the RMA is an asymptotic pseudotrajectory of the semiflow induced by $F$ almost surely. \label{RM-converge-thm}
}

The semiflows we will consider are \emph{gradient flows}. A gradient flow is a semiflow $\Phi$ for which
there exists a potential function $G$ such that $\frac{\partial}{\partial t} G(\Phi(t,\bbz)) \le 0$.
Borkar~\cite{Borkar2008} presented a theorem which is convenient for our usages.

\boxthm{[Borkar~\cite{Borkar2008}, Corollary 3] \label{thm:sa-borkar} Let $\Phi$ be the semiflow on $\rr^m$, induced by a smooth vector field $F:\rr^m \ra \rr^m$.
Suppose there is a continuously differentiable function $G:\rr^m \rightarrow \rr$ such that
\begin{enumerate}
\item $\lim_{\|\bbz\|\rightarrow \infty} G(\bbz) = +\infty$;
\item the set of minimum points of $G$, denoted by $Z^*$, is non-empty; and
\item $\innerProd{\nabla G(\bbz)}{F(\bbz)} \le 0$ for all $\bbz$, with equality holds if and only if $\bbz\in Z^*$,
\end{enumerate}
Then the corresponding RMA with step-sizes satisfying the conditions in \Cref{thm:sa-benaim} converges to a non-empty subset of $G^*$ almost surely.}

\subsection{Formulation of Stochastic T-O Market Dynamic as Robbin-Monro Algorithm}\label{Formulation of the Robbin-Monro Algorithm}

The proofs of \Cref{thm-homo-convergence,thm:convergence-hetero} follow the same approach.
First, we follow Maldonado et al.~\cite{maldonado2018popularity}'s approach to rewrite the stochastic dynamics in T-O markets as RMA.
We present the following proposition which deals with general settings, which may be of wider interest.

\boxprop{\label{prop:rma-general} There are $m$ possible outcomes. Suppose there is a discrete-time stochastic process,
where at each integer $t\ge 1$, exactly one of the $m$ outcomes occurs.
For each $t\ge 0$, let $\bbn^t = (n^t_1,n^t_2,\ldots,n^t_m)$ denote the vector that counts the number of occurrences of each outcome on or before time $t$.
We assume that $n^0_j \ge 0$ and $N_0 := \sum_{j=1}^m n^0_j \ge 1$, i.e., there is some occurrence of outcomes before the stochastic process begins.
For $t\ge 0$, let $\bbz^t = (\frac{n^t_1}{N_0+t},\frac{n^t_2}{N_0+t},\ldots,\frac{n^t_m}{N_0+t})\in \Delta^m$
denote the vector that represents the fraction of past occurrences of each outcome.
Suppose there is a function $F:\Delta^m \ra \Delta^m$, such that the $j$-th component of $F(\bbz^{t-1})$ is the probability that outcome $j$ occurs at time $t$.
Then $\{\bbz^t\}_{t\in\nn}$ can be written as a RMA of the form
\[
\bbz^t ~=~ \bbz^{t-1} ~+~ \frac{1}{N_0 + t} \cdot \left[(F(\bbz^{t-1}) - \bbz^{t-1}) + \bbu^t\right]~.\vuppp\vuppp
\]
}

\begin{pf}
Suppose $z_j^{t-1} = \frac{a}{N_0+t-1}$ for some integer $a\ge 0$.
Note that $z_j^t - z_j^{t-1}$ is a random variable which has value $\frac{a+1}{N_0+t} - \frac{a}{N_0+t-1} = \frac{N_0+t-1-a}{(N_0+t)(N_0+t-1)}$ with probability $[F(\bbz^{t-1})]_j$,
and it has value $\frac{a}{N_0+t} - \frac{a}{N_0+t-1} = \frac{-a}{(N_0+t)(N_0+t-1)}$ otherwise. Thus, the expected value of $z_j^t - z_j^{t-1}$ is
\[
\frac{[F(\bbz^{t-1})]_j \cdot (N_0+t-1-a) + (1-[F(\bbz^{t-1})]_j) \cdot (-a)}{(N_0+t)(N_0+t-1)} = \frac{1}{N_0+t} \cdot \left( [F(\bbz^{t-1})]_j - z_j^{t-1} \right)~.
\]
The proposition follows.
\end{pf}

By using \Cref{thm:sa-borkar} with the above proposition, we have the following lemma.

\boxlem{\label{lem:rma-gradientflow} Consider the stochastic process described in \Cref{prop:rma-general}.
If the semiflow induced by the ODE system $\dot{\bbz} = F(\bbz)-\bbz$ is a gradient flow in $\Delta^m$ with potential function $G$,
then $\{\bbz^t\}_{t\in\nn}$ of the stochastic process converges to a subset of $\argmin_{\bbz\in \Delta^m} G(\bbz)$ almost surely.}

Now we move back to our problem. Let $\bbb^t$ denote a vector in $\Delta^{(|\calU|\times |\calI| + 1)}$, such that for each user type $i\in \calU$ and item $j\in \calI$,
$b_{ij}^t$ is the fraction of users of user type $i$ who tried and purchased item $j$ on or before time $t$.
The extra dimension in $\bbb^t$ concerns the remaining users, i.e., the fraction of users who did not purchase an item on or before time $t$.

Recall from \Cref{sec:model} that given $\bbb^{t-1}$, the probability that the next user is of type $i$ and she tries and purchases item $j$ is
$w_i q_{ij} \cdot \frac{v_{ij} (b^{t-1}_j)^{r_i}}{\sum_{k=1}^{|\calI|} v_{ik} (b^{t-1}_k)^{r_i}}$, which is exactly the RHS of the dynamic \eqref{cont dynamic}.
We let $[F(\bbb^{t-1})]_{ij}$ denote the above probability.
Also, \Cref{Continuous_dynamics_as_mirror_descent} states that \eqref{cont dynamic} is indeed a mirror descent update rule for a convex optimization problem.
Thus, we expect the semiflow $\dot{\bbb} = F(\bbb)-\bbb$ to be a gradient flow.
We will formally prove this in the next subsection, then we are done by applying \Cref{lem:rma-gradientflow}.

\subsection{From Discrete-Time Mirror Descent to Continuous-Time Gradient Flow}\label{sec:supp-shorter-proof}

In this subsection, we will prove \Cref{lem:discrete-time-to-cont-time} below which is for general conversion
of discrete-time mirror descent update rules to continuous-time gradient flow.
The lemma can be applied to dynamic \eqref{cont dynamic} to conclude the above-mentioned semiflow $\dot{\bbb} = F(\bbb)-\bbb$ is a gradient flow.

\boxdef{Let $C$ be a compact and convex set. Given a differentiable convex function $h$ with convex domain $C$,
the Bregman divergence generated by $h$, which is denoted by $d_h$ is defined as
\[
d_h(\z,\y) = h(\z) - h(\y) - \langle\nabla h(\y), \z - \y \rangle.
\]
where $\z \in C$, $\y \in \text{rint}(C) = \{\y\in C: \forall \x\in C,~\exists\lambda>1 \text{ such that } \lambda \y+ (1-\lambda)\x \in C\}$.
}

We assume that the function $h$ is strictly convex. Consequently, $d_h(\cdot,\cdot)$ is strictly convex in its first parameter, and $d_h(\z,\y) = 0$ if and only if $\z = \y$.

\boxdef{A function $G$ is $L$-Bregman-convex with respect to the Bregman divergence $d_h$ if
for any $\y \in \text{rint}(C)$ and $\z\in C$,
\[
G(\z)  ~\leq~ G(\y) + \langle\nabla G(\y), \z-\y\rangle + L \cdot d_h(\z,\y).\vuppp\vuppp
\]
}

Given an $L$-Bregman-convex function $f$ with respect to the Bregman divergence $d_h$,
the mirror descent rule with respect to the Bregman divergence $d_h$ is given by $\z^{t} = F(\z^{t-1})$, where
\begin{equation}
F(\z^{t-1}) = \argmin_{\z\in C} \left\{G(\z^{t-1}) + \langle\nabla G(\z^{t-1}), \z - \z^{t-1}\rangle +L \cdot d_h(\z,\z^{t-1}) \right\}. \label{MDMinimizer-again}
\end{equation}

\boxlem{\label{lem:discrete-time-to-cont-time} Let $C$ be a compact and convex set, and $h$ be a strictly convex function on $C$.
Suppose $G:C\ra\rr$ is $L$-Bregman-convex with respect to the Bregman divergence $d_h$.
Let $F$ be the mirror descent update rule \eqref{MDMinimizer-again}. Then $\dot{\bbz} = F(\bbz)-\bbz$ is a gradient flow.}

To prove the above lemma, we use the following convergence result of mirror descent from Chen and Teboulle~\cite{CT93}. Let $Z^* = \argmin_{\bbz\in C} G(\bbz)$.

\boxthm{[Chen and Teboulle~\cite{CT93}] \label{CT93}
Suppose $G$ is an $L$-Bregman-convex function with respect to $d_h$, and $\z^t$ is the point reached after $t$ applications of
the mirror descent update rule $\z^{t} = F(\z^{t-1})$. Then for all $t\ge 1$ and $\bbz^*\in Z^*$,
\[
G(\z^t) - G(\z^*) ~\le~ \frac{L\cdot d_h(\z^*, \z^0 )}{t}.\vuppp\vuppp
\]
}

We also need the following lemma which shows each mirror descent update strictly reduces the value of $G$, unless it is already in the set of minimum points.

\boxlem{\label{lem:strict-decrease} (a) If $\z \notin Z^*$, then $F(\z) \neq \z$ and $G(F(\z)) < G(\z)$. (b) If $\z \in Z^*$, then $F(\z) = \z$.}

\begin{pf}
First, we show that if $\z\notin Z^*$, then $F(\z) \neq \z$.
Suppose the contrary, i.e., $\z$ is a fixed point of the update rule $\z^{t} = F(\z^{t-1})$.
By setting $\z^0 = \z$, \Cref{CT93} implies that $G(\z) - G(\z^*) \le \frac{L\cdot d_h(\z^*, \z )}{t}$ for any $t\ge 1$.
But the LHS does not depend on $t$, this is possible only when $G(\z) - G(\z^*) = 0$ and hence $\z \in Z^*$, a contradiction.

Let $\hat{G}(\y) = G(\z) + \langle\nabla G(\z), \y - \z\rangle +L \cdot d_h(\y,\z)$.
By the definition of $F$, we have $\hat{G}(F(\z)) \le \hat{G}(\z)$.
Since $d_h(\cdot,\cdot)$ is strictly convex in its first parameter, $\hat{G}$ is strictly convex.
Since $F(\z) \neq \z$ and $\hat{G}$ is strictly convex, for any $\z'$ on the line segment between $\z$ and $F(\z)$,
$\hat{G}(\z') < \hat{G}(\z)$ and $\hat{G}(\z') < \hat{G}(F(\z))$.
By the definition of $F$ again, we have $\hat{G}(F(\z)) \le \hat{G}(\z')$.
Since $G$ is $L$-Bregman-convex with respect to $d_h$, $G(F(\z)) \le \hat{G}(F(\z))$.
Combining all the above inequalities gives
\[
G(F(\z)) ~\le~ \hat{G}(F(\z)) ~\le~ \hat{G}(\z') ~<~ \hat{G}(\z) ~=~ G(\z)~.
\]

The last paragraph shows that if $F(\z) \neq \z$ then $G(F(\z)) < G(\z)$. However, if $\z \in Z^*$, it is impossible that for $G(z)$ to be strictly larger than $G(F(\z))$.
Thus, if $\z \in Z^*$ then $F(\z) = \z$.
\end{pf}

\begin{pfof}{\Cref{lem:discrete-time-to-cont-time}}
We want to show $\innerProd{\nabla G(\z)}{F(\z) - \z} \le 0$, with equality holds if and only if $\z\in Z^*$.
For any $\bbz\notin Z^*$ consider $G$ restricted on the line segment between $\z$ and $F(\z)$.
By \Cref{lem:strict-decrease}, $F(\z) < \z$. Moreover, $G$ is convex. Thus, for any $\lambda \in [0,1]$,
\[G((1-\lambda)\cdot \z + \lambda \cdot F(\z)) - G(\z) \le \lambda \cdot \left[ G(F(\z)) - G(\z) \right]\].
Dividing both sides of the above inequality by $\lambda$, and then taking $\lambda \searrow 0$,
the LHS becomes $\innerProd{\nabla G(\z)}{F(\z) - \z}$, while the RHS is $G(F(\z)) - G(\z)$ which is strictly negative.

If $\z\in Z^*$, \Cref{lem:strict-decrease} implies that $F(\z) = \z$, and hence $\innerProd{\nabla G(\z)}{F(\z) - \z} = 0$.
\end{pfof}
\section{Additional simulation results}\label{app:simulations}

\begin{figure*}[h!]
\centering
\begin{minipage}{0.52\textwidth}
\centering
\includegraphics[width=1.\textwidth]{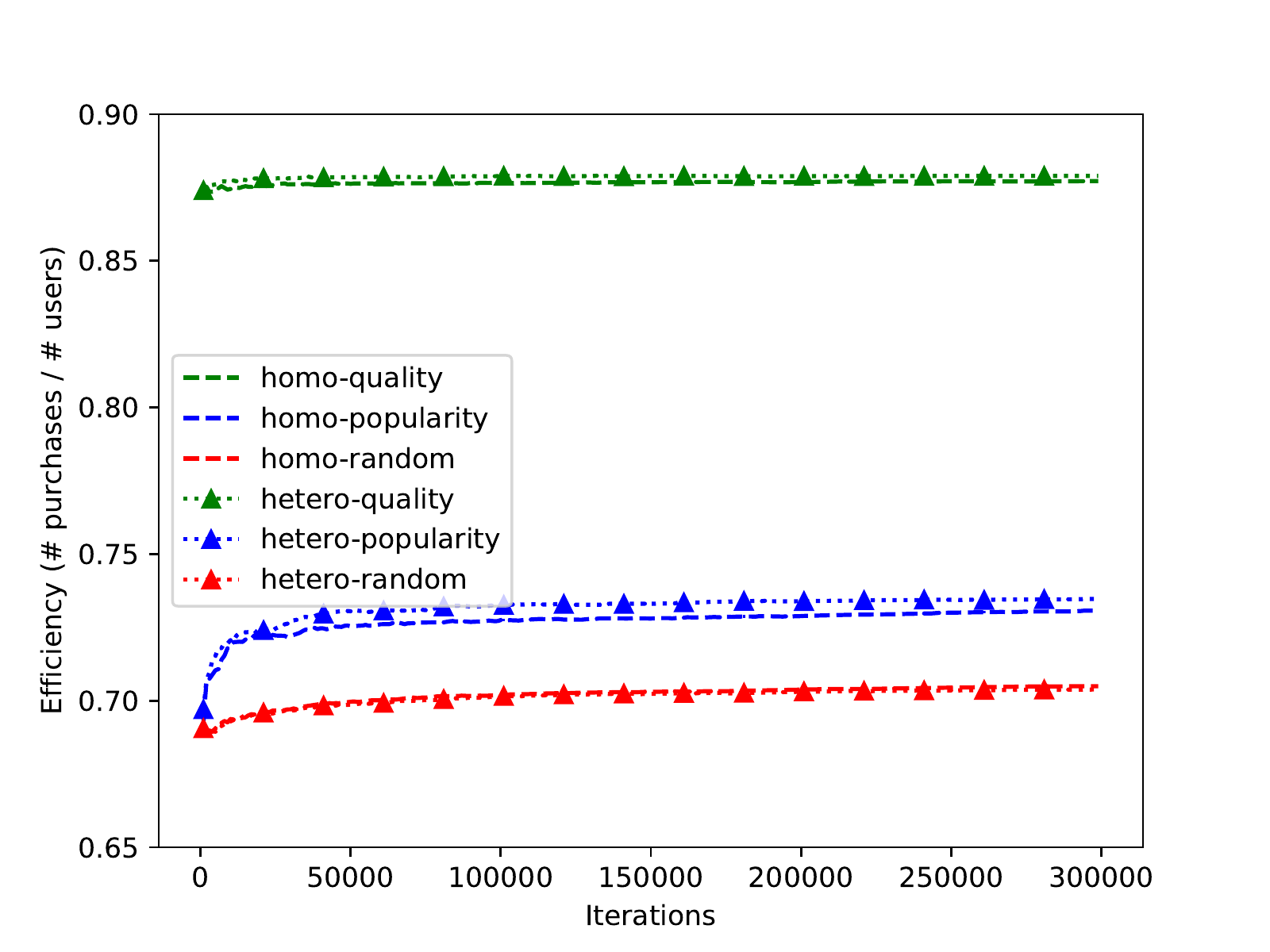}%
\end{minipage}%
\begin{minipage}{0.52\textwidth}
\centering
\includegraphics[width=1.\textwidth]{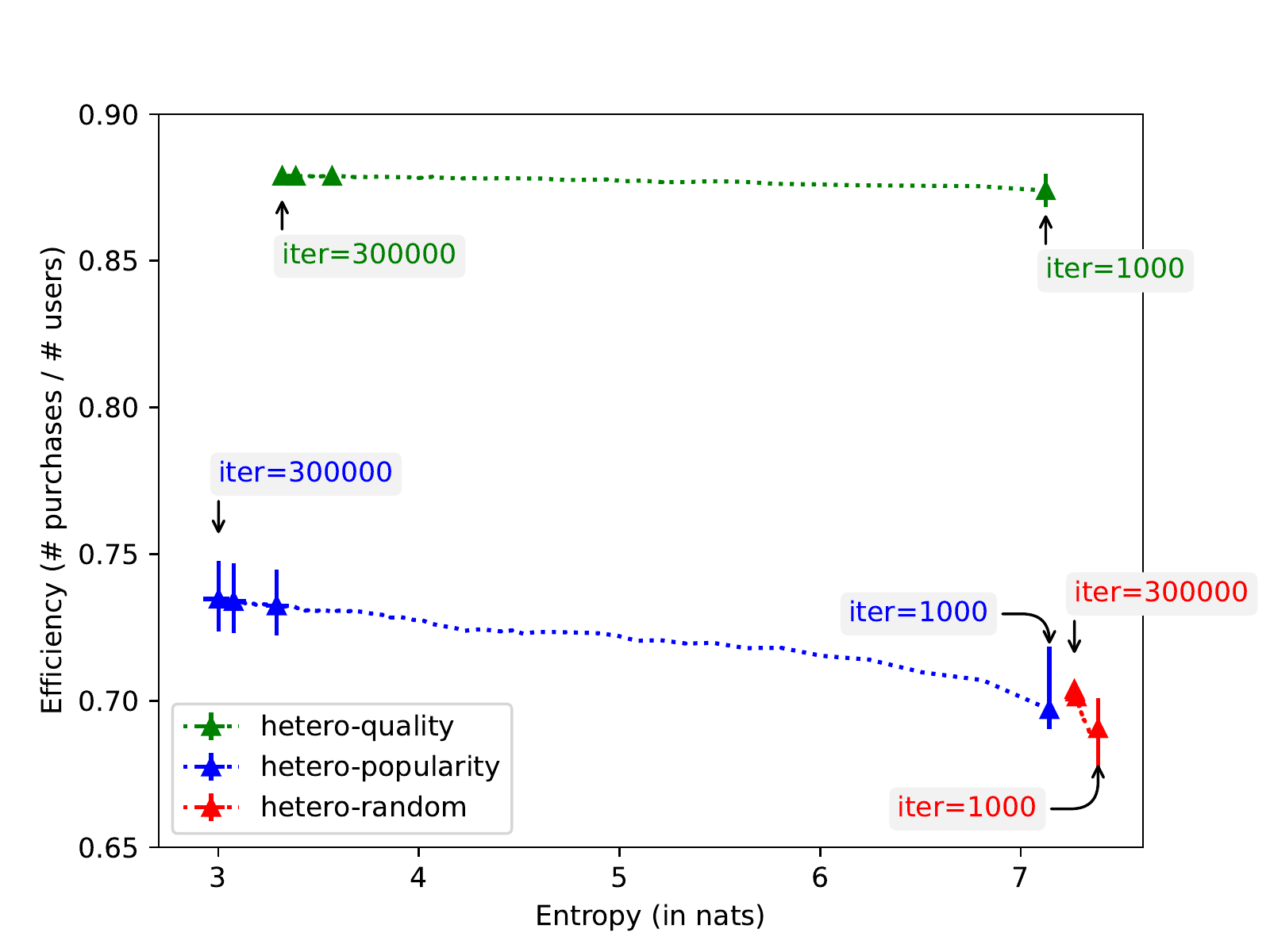}%
\end{minipage}%
\caption{Simulation results on MovieLens dataset with 50 user groups.}
\label{fig:movielens50}
\end{figure*}

\begin{figure*}[h!]
\centering
\begin{minipage}{0.52\textwidth}
\centering
\includegraphics[width=1.\textwidth]{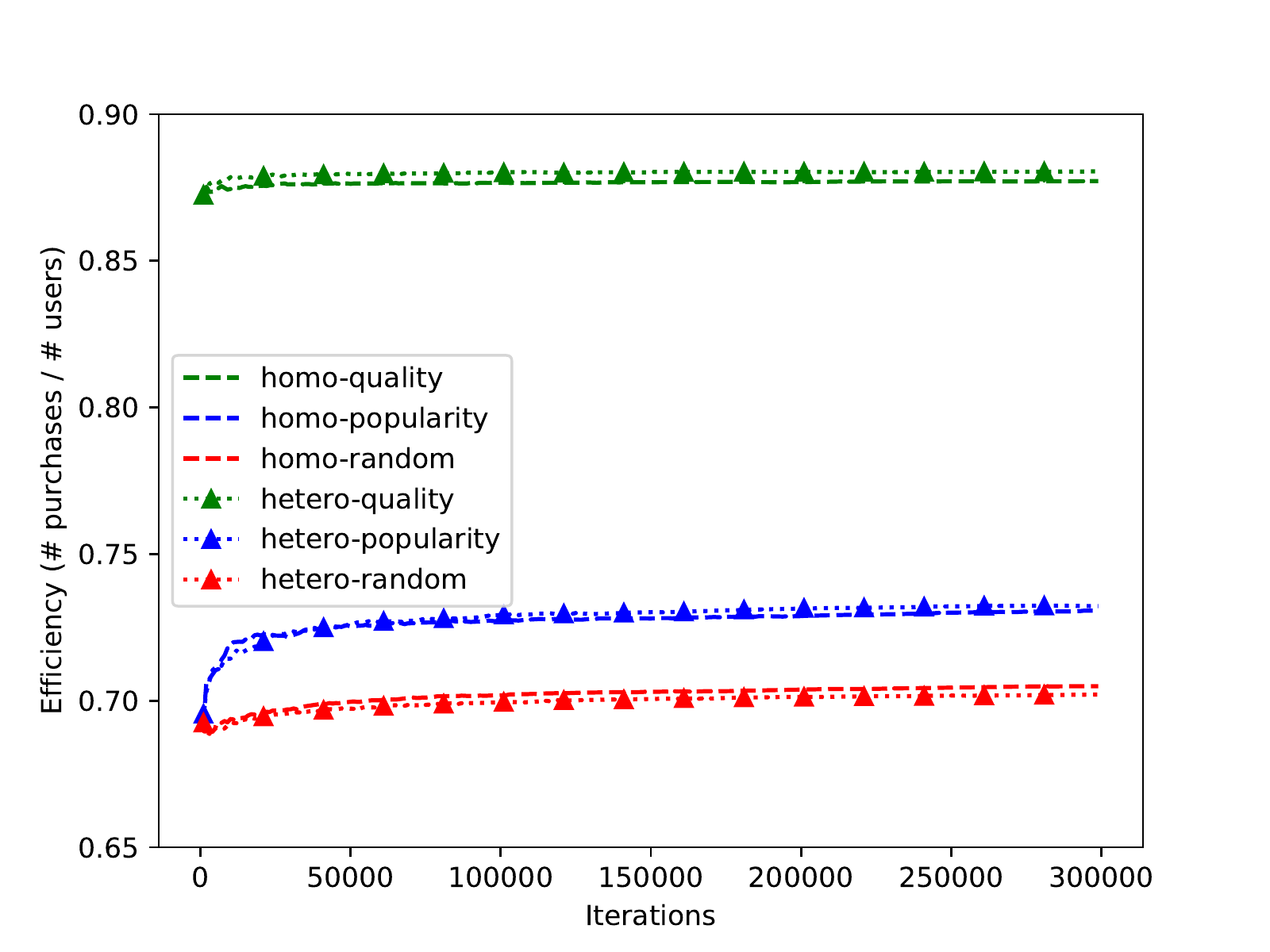}%
\end{minipage}%
\begin{minipage}{0.52\textwidth}
\centering
\includegraphics[width=1.\textwidth]{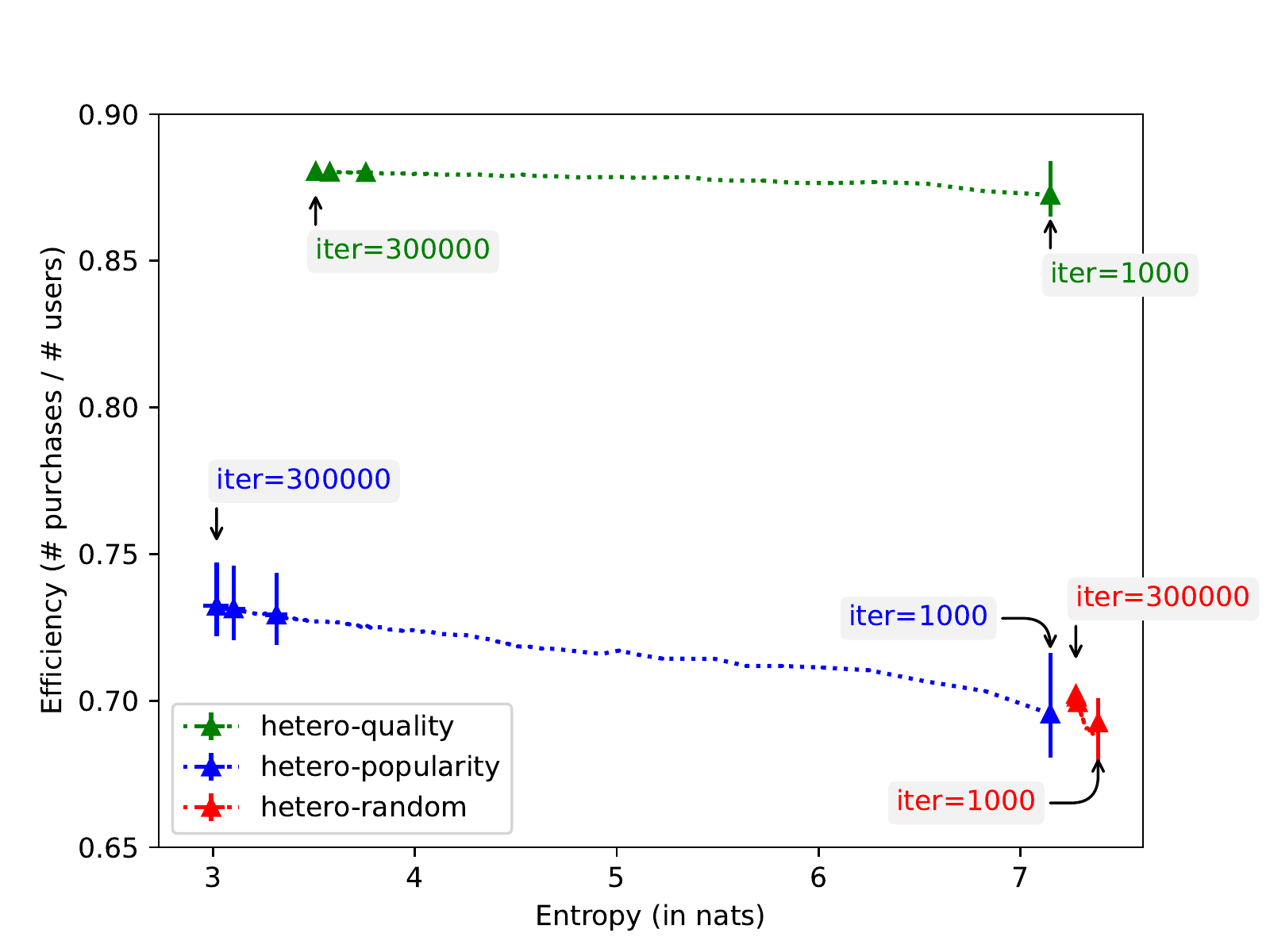}%
\end{minipage}%
\caption{Simulation results on MovieLens dataset with 200 user groups.}
\label{fig:movielens200}
\end{figure*}

\begin{figure*}[h!]
\centering
\begin{minipage}{0.52\textwidth}
\centering
\includegraphics[width=1.\textwidth]{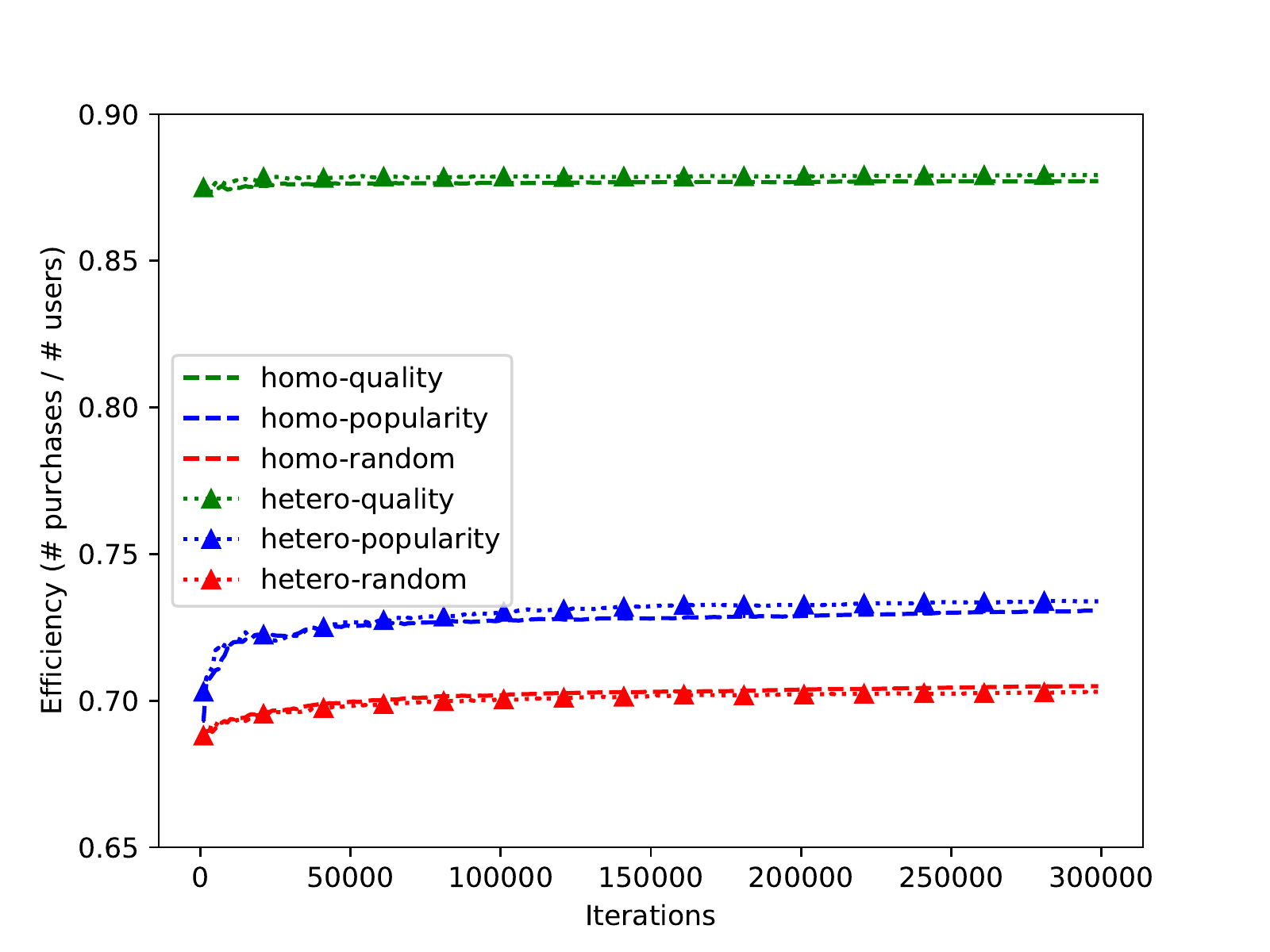}%
\end{minipage}%
\begin{minipage}{0.52\textwidth}
\centering
\includegraphics[width=1.\textwidth]{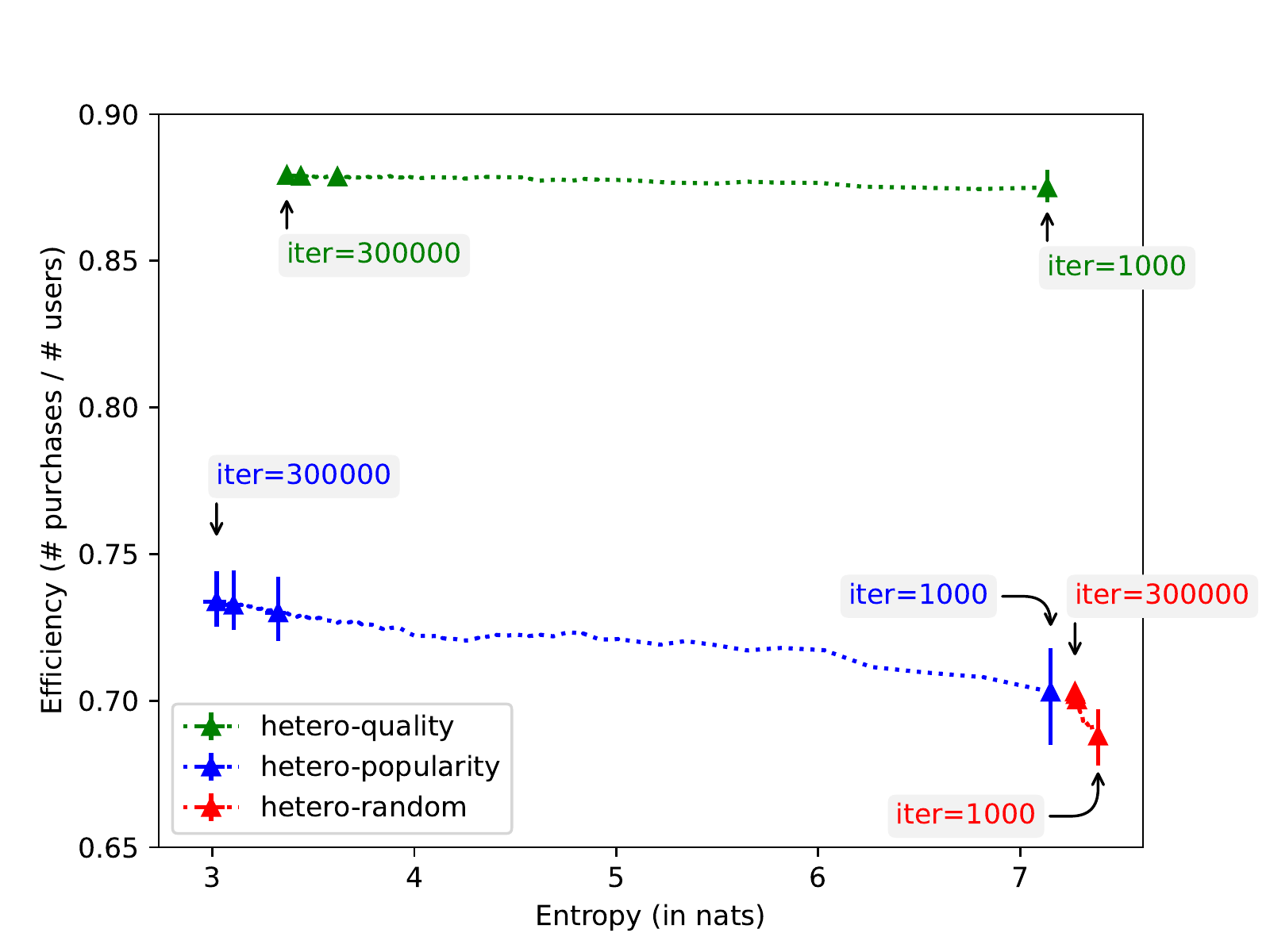}%
\end{minipage}%
\caption{Simulation results on MovieLens dataset with users only choosing among unseen movies.}
\label{fig:100groups_unseen_only}
\end{figure*}

We provide additional simulation results in other settings. \Cref{fig:movielens50} and \Cref{fig:movielens200} provide the simulation results with 50 and 200 user groups obtained by K-means clustering. \Cref{fig:100groups_unseen_only} is the setting where the users could only choose among the movies that they have not seen (i.e. the choices denoted by user-item pairs could only be in the set $\Tilde{O}$). Compared to the simulation result we have already analysed in the main text, there are no outstanding differences in the observations.

\end{document}